\documentclass[12pt]{article}
\usepackage{jheppub}

\pdfoutput=1

\usepackage{amsmath,bbm,array,amsfonts,graphicx,wrapfig,lscape,float,mathtools,multirow,longtable}
\usepackage[dvipsnames]{xcolor}

\newcommand{\be}{\begin{equation}}
\newcommand{\ee}{\end{equation}}
\newcommand{\beq}{\begin{equation}}
\newcommand{\beql}[1]{\begin{equation}\label{#1}}
\newcommand{\eeq}{\end{equation}}
\newcommand{\ba}{\begin{array}}
\newcommand{\ea}{\end{array}}
\newcommand{\bea}{\begin{eqnarray}}
\newcommand{\beal}[1]{\begin{eqnarray}\label{#1}}
\newcommand{\eea}{\end{eqnarray}}
\newcommand{\ben}{\begin{enumerate}}
\newcommand{\een}{\end{enumerate}}
\newcommand{\bean}{\begin{eqnarray*}}
\newcommand{\eean}{\end{eqnarray*}}
\newcommand{\eref}[1]{(\ref{#1})}
\newcommand{\sref}[1]{\S\ref{#1}}

\newcommand{\nn}{\nonumber}

\newcommand{\fref}[1]{Figure \ref{#1}}
\newcommand{\btab}[1]{\begin{tabular}{#1}}
\newcommand{\etab}{\end{tabular}}

\def \Black {\pdfsetcolor{0 0 0 1}}

\newcommand{\comment}[1]{}

\newcommand{\qed}{\nobreak \ifvmode \relax \else
      \ifdim\lastskip<1.5em \hskip-\lastskip
      \hskip1.5em plus0em minus0.5em \fi \nobreak
      \vrule height0.75em width0.5em depth0.25em\fi}

\newcommand{\Tr}{\text{Tr}}

\definecolor{darkspringgreen}{rgb}{0.09, 0.45, 0.27}
\definecolor{forestgreen}{rgb}{0.13, 0.55, 0.13}

\usepackage{array}
\newcolumntype{C}[1]{>{\centering\let\newline\\\arraybackslash\hspace{0pt}}m{#1}}

\title{Quadrality for Supersymmetric Matrix Models} 
\author[a,b]{Sebasti\'an Franco,} 
\author[c,d,e]{Sangmin Lee,}
\author[f]{Rak-Kyeong Seong,}
\author[g]{Cumrun Vafa}

\affiliation[a]{
Physics Department, The City College of the CUNY \\
160 Convent Avenue, New York, NY 10031, USA}

\affiliation[b]{The Graduate School and University Center, The City University of New York  \\
365 Fifth Avenue, New York NY 10016, USA}

\affiliation[c]{
Center for Theoretical Physics, Seoul National University, Seoul 08826, Korea
}

\affiliation[d]{
Department of Physics and Astronomy, Seoul National University, Seoul 08826, Korea
}

\affiliation[e]{
College of Liberal Studies, Seoul National University, Seoul 08826, Korea
}

\affiliation[f]{
Department of Physics and Astronomy, Uppsala University, SE-751 08 Uppsala, Sweden
}

\affiliation[g]{Jefferson Physical Laboratory, Harvard University, Cambridge, MA 02138, USA}

\emailAdd{sfranco@ccny.cuny.edu}
\emailAdd{sangmin@snu.ac.kr}
\emailAdd{rakkyeongseong@gmail.com}
\emailAdd{vafa@physics.harvard.edu}

\preprint{
\begin{flushright}
CCNY-HEP-16-10 \\
SNUTP16-007 \\
UUITP-31/16
\end{flushright}
}

\abstract{We introduce a new duality for $\mathcal{N}=1$ supersymmetric gauged matrix models. This $0d$ duality is an order 4 symmetry, namely an equivalence between four different theories, hence we call it Quadrality. Our proposal is motivated by mirror symmetry, but is not restricted to theories with a D-brane realization and holds for general $\mathcal{N}=1$ matrix models. We present various checks of the proposal, including the matching of: global symmetries, anomalies, deformations and the chiral ring. We also consider quivers and the corresponding quadrality networks. Finally, we initiate the study of matrix models that arise on the worldvolume of D(-1)-branes probing toric Calabi-Yau 5-folds.
}

\begin{document}

\maketitle

%=================================================================
\section{Introduction}
%=================================================================

Duality, the equivalence between seemingly different theories, is one of the most fascinating phenomena in quantum field theory. Some of the best-understood examples of dualities involve supersymmetry, since it provides an enhanced control of the theories. 

Seiberg duality, an equivalence between two $4d$ $\mathcal{N}=1$ gauge theories in the IR limit, is the prototypical example of a supersymmetric duality \cite{Seiberg:1994pq}. More recently, Gadde, Gukov and Putrov (GGP) discovered that $2d$ $\mathcal{N}=(0,2)$ gauge theories exhibit triality \cite{Gadde:2013lxa}. This is an IR equivalence, analogous to Seiberg duality, but of order $3$. This means that triality relates three different theories and that the original theory is recovered after three consecutive transformations. The existence of triality makes it clear that the space of dualities is far richer than naively suspected and that, in general, it involves order $n$ dualities.

The realization of quantum field theories in terms of branes is an extremely fruitful approach for understanding and uncovering dualities. In a recent paper \cite{Franco:2016qxh}, building on \cite{Cachazo:2001sg}, we showed that mirror symmetry provides a geometric unification of dualities in different dimensions. Mirror symmetry naturally explains why $(10 -2n)$-dimensional quantum field theories, which are associated with CY $n$-folds, exhibit duality symmetries of order $n-1$. Our work not only explained $4d$ Seiberg duality and $2d$ GGP triality in a unified framework, but also led us to conjecture an order $4$ duality for $\mathcal{N}=1$ gauged matrix models in $0d$, which we called {\it quadrality}.\footnote{We can regard matrix models as quantum field theories in $0d$.} The main goal of this paper is to introduce quadrality and to perform various checks of the proposal. 

Even though we use mirror symmetry to motivate quadrality, our claim is much stronger: we postulate it is a property of general $\mathcal{N}=1$ gauged matrix models. Quadrality applies to matrix models with potentials, which are ubiquitous in physics. $\mathcal{N}=1$ supersymmetry can be regarded as an appropriate dressing by fermions that allows for additional control of the theory. 

The organization of this paper is as follows. Section \sref{sec:super-matrix} introduces the basics of $\mathcal{N}=1$ gauged matrix models. Section \sref{section_geometric_motivation} motivates quadrality by considering the $0d$ $\mathcal{N}=1$ theories on D(-1)-branes probing toric Calabi-Yau (CY) 5-folds and mirror symmetry. Section \sref{section_quadrality} introduces quadrality for general $\mathcal{N}=1$ gauged matrix models and elaborates on its physical meaning. Various checks of the proposal are presented in section \sref{section_checks}. Section \sref{section_quadrality_networks} discusses how theories connected by sequences of quadrality transformations can be organized into quadrality networks. Section \sref{section_D-brane_examples} considers theories on the worldvolume of D(-1)-branes probing toric CY 5-folds. We also show, in the local $\mathbb{CP}^4$ example, how the chiral ring of the gauge theory reproduces the coordinate ring of the probed geometry and remains invariant under quadrality. We present our conclusions in section \sref{section_conclusions}. In two appendices we review the application of Hilbert series to the chiral rings of gauge theories on D-branes.

%=================================================================
\section{Supersymmetric Gauged Matrix Models} \label{sec:super-matrix}
%=================================================================

In this paper we study $\mathcal{N}=1$ supersymmetric gauged matrix models. This section presents the basic properties of these theories. Despite the fact that the theories carry only one supercharge $Q$, some useful concepts in supersymmetric theories such as holomorphy and $R$-symmetry are still valid. The main reason is that 
in matrix models, unlike in quantum mechanics or QFT, we do not have to consider Hermitian conjugates of operators. In the matrix integral, all Fermi fields are regarded as independent holomorphic variables without Hermitian conjugates. 

The basic multiplets of the supersymmetric matrix models are as follows: 
\begin{enumerate}

\item
Gaugino multiplet $V_\alpha$:
\begin{align}
\{Q, \chi_\alpha \} = D_\alpha \,,
\quad
[Q, D_\alpha] = 0 \,. 
\label{algebra_gaugino}
\end{align}

\item 
Chiral multiplet $X_i$:
\begin{align}
[Q, \phi_i] = \psi_i \,,
\quad
\{Q, \psi_i\} = 0 \,,
\quad
[Q, \bar{\phi}_i ] = 0 \,. 
\label{algebra_chiral}
\end{align}

\item 
Fermi multiplet $\Lambda_a$: 
\begin{align}
\{Q, \lambda_a \} = G_a \,,
\quad
[Q, G_a] = 0 \,, 
\quad
\{Q, \bar{\lambda}^a \} = \overline{G}^a \,,
\quad
[Q, \overline{G}^a] = 0 \,. 
\label{algebra_Fermi}
\end{align}

\end{enumerate}

\noindent
Each line presents the components of a supermultiplet and how they transform under the action of the supercharge. We have adopted a notation that resembles that of $2d$ $(0,2)$ gauge theories.\footnote{This resemblance may cause some confusion. See \sref{sec:dim-red} for precise relations.} The superalgebra $Q^2=0$ holds trivially in all multiplets. As far as the superalgebra is concerned $(\phi,\psi)$ are independent of $\bar{\phi}$ and $(\lambda,G)$ are independent of $(\bar{\lambda},\overline{G})$. However, in view of the structure of interaction terms, it is convenient to regard $(\phi,\psi;\bar{\phi})$ as a single multiplet and $(\lambda,G; \bar{\lambda},\overline{G})$ as another one. 

We now describe the general structure of $\mathcal{N}=1$ gauged matrix models, focusing on the abelian case. The non-abelian extension is straightforward. At least locally in field space, both the $D$-term and $J$-term contributions are $Q$-exact, namely
\begin{align}
S_D = \{Q , \Sigma_D \}  \,,
\quad
S_J = \{Q , \Sigma_J \} \,.
\end{align}
The ``Fermionic Lagrangian" terms are given by 
\begin{align}
\begin{split}
 \Sigma_D &= \sum_{\alpha=1}^G \chi_\alpha \left( -\frac{1}{2}D_\alpha + \sum_{i=1}^C q_{\alpha i} \phi_i \bar{\phi}_i -t_\alpha \right)   \,,
\\
\Sigma_J &= \sum_{a=1}^F 
\left( \bar{\lambda}^a (\bar{J}_a(\bar{\phi})-G_a) + \lambda_a J^a(\phi) \right) \,.
\end{split}
\end{align}
Expanding in components and integrating out auxiliary fields, we obtain
\begin{align}
\begin{split}
S_D &=  \sum_\alpha \left[ -\frac{1}{2} D_\alpha^2 + D_\alpha \left(\sum_i q_{\alpha i} \phi_i \bar{\phi}_i -t_\alpha \right)- \chi_\alpha \sum_i q_{\alpha i} \psi_i \bar{\phi}_i \right] 
\\
&\simeq
\sum_\alpha \left[\frac{1}{2} \left(\sum_i q_{\alpha i} \phi_i \bar{\phi}_i -t_\alpha \right)^2 - \chi_\alpha\sum_i q_{\alpha i} \psi_i \bar{\phi}_i \right] \,,
\\
S_J &=  \sum_{a} 
\left[ \overline{G}^a (\bar{J}_a(\bar{\phi})-G_a) + G_a J^a(\phi) + \sum_i \lambda_a \frac{\partial J^a}{\partial \phi_i} \psi_i \right] 
\\
&\simeq
\sum_a \left[ \bar{J}_a(\bar{\phi}) J^a(\phi) + \sum_i \lambda_a \frac{\partial J^a}{\partial \phi_i} \psi_i \right] \,.
\end{split}
\end{align}
We have fixed the overall sign of $S_D$ and $S_J$ such that their bosonic terms are positive definite when 
$\phi$ and $\bar{\phi}$ satisfy the reality condition $(\phi_i)^* = \bar{\phi}_i$. This condition is necessary in order for the matrix integral $\int \mathcal{D}(\phi,\psi,\lambda,\chi) e^{-S(\phi,\psi,\lambda,\chi)}$ to converge. 
In contrast, as mentioned earlier, we do not impose any reality condition for fermions.

%=================================================================
\subsection{Novel Interaction Terms}
%=================================================================

$\mathcal{N}=1$ matrix models allow for yet another type of interaction, which we call $H$-terms: 
\begin{align}
S_H = \frac{1}{2} \sum_{a,b=1}^F \overline{H}^{ab}(\bar{\phi}) \lambda_a \lambda_b 
\,,
\end{align}
where $\overline{H}_{ab} = - \overline{H}_{ba}$ are antiholomorphic functions that depend exclusively on the $\bar{\phi}$'s.

Unlike $S_D$ and $S_J$, $S_H$ is {\em not} locally exact. Using the fact that $\{Q, \lambda_a \} = G_a$ and that $G_a = \bar{J}_a(\bar{\phi})$ {\em on-shell}, we obtain
\begin{align}
\{ Q, S_H \} = \sum_{a,b} \overline{H}^{ab} (\bar{\phi}) \bar{J}_a(\bar{\phi}) \lambda_b \,.
\end{align}
Since every $\lambda_a$ is independent, $S_H$ is supersymmetric if and only if 
\begin{align}
\sum_{b=1}^F \overline{H}^{ab} \bar{J}_b = 0 \quad 
\mbox{for each }a\,.
\label{H-constraint}
\end{align}
We refer to this condition as the $H$-constraint.

%=================================================================
\paragraph{Mass Terms.}
%=================================================================

\begin{figure}[H]
	\centering
	\includegraphics[height=4cm]{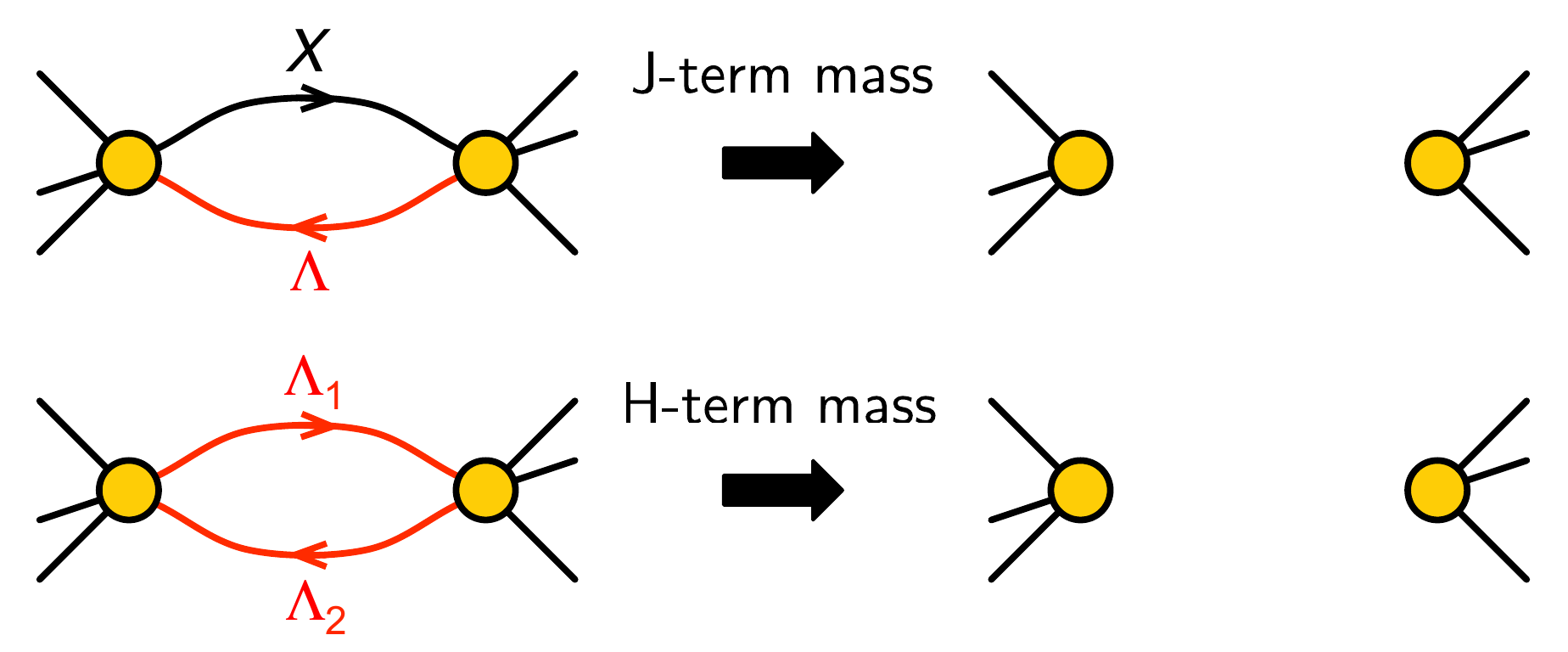}
\caption{Two types of mass terms in an $\mathcal{N}=1$ matrix model.}
	\label{pair-cancel}
\end{figure}

The interactions that we just described give rise to two types of mass terms as summarized in \fref{pair-cancel}: a chiral-Fermi mass via a $J$-term and a Fermi-Fermi mass via an $H$-term. They take the following form
\beq
\begin{array}{lrcl}
\mbox{$J$-term mass:} \ \ \ & J_\Lambda & = & m \, \phi_X \, , \\[.1cm]
\mbox{$H$-term mass:} \ \ \ & H_{\Lambda_1,\Lambda_2} & = & m \, .
\end{array}
\eeq

%=================================================================
%=================================================================

%=================================================================
\paragraph{Superspace Notation and Products.}
%=================================================================

With $0d$ $\mathcal{N}=1$ SUSY, it is possible to introduce a Grassmann coordinate $\theta$ and package the different components into superfields. For the matter supermultiplets we have
\beq
\begin{array}{ccl}
X& = & \phi+\theta \, \psi\, , \ \ \ \ \ \ \overline{X}=\bar{\phi} \, , \\[.1cm]
\Lambda & = & \lambda+\theta \, G \, .
\end{array}
\eeq
The basic distinction between them resides in the commutation properties of their lowest components. Let us now consider products of superfields.

\begin{itemize}
%=================================================================
\item \underline{$\mbox{chiral} \times \mbox{chiral} = \mbox{chiral}$}
%=================================================================

In detail, we have
\beq
X_1 X_2 = \phi_1 \phi_2 + \theta (\phi_1 \psi_2+\psi_1 \phi_2) \, , \ \ \ \ \ \  \overline{X}_1 \overline{X}_2 = \bar{\phi}_1 \bar{\phi}_2 \, .
\label{product_1}
\eeq
It is straightforward to verify that the result is a chiral field, since the two components transform under the action of $Q$ as the $\phi$ and $\psi$ components in \eref{algebra_chiral}.

%=================================================================
\item \underline{$\mbox{chiral} \times \mbox{Fermi} = \mbox{Fermi}$} 
%=================================================================

The product becomes
\beq
X_1  \Lambda_2 = \phi_1 \lambda_2 + \theta (\phi_1 G_2 + \psi_1 \lambda_2) \, ,
\label{product_2}
\eeq
which corresponds to a Fermi field, since the components behave as $\lambda$ and $G$ in \eref{algebra_Fermi}.

%=================================================================
\item \underline{$\overline{\mbox{chiral}} \times \mbox{Fermi} = \mbox{Fermi}$}
%=================================================================

In this case, we have
\beq
\overline{X}_1  \Lambda_2 = \bar{\phi}_1 \lambda_2 + \theta \bar{\phi}_1 G_2 \, ,
\label{product_3}
\eeq
which, following \eref{algebra_Fermi}, is also a Fermi field.

%=================================================================
\item \underline{$\mbox{Fermi} \times \mbox{Fermi}$}
%=================================================================

Finally, let us consider the product of two Fermi fields. We have
\beq
\Lambda_1 \Lambda_2 = \lambda_1 \lambda_2 + \theta (G_1 \lambda_2-\lambda_1 G_2)  \, .
\label{product_4}
\eeq
Comparing with \eref{algebra_chiral}, this naively looks like a chiral field. However, the $\bar{\phi}$ component of this would-be chiral field is automatically zero, hence absent. We conclude this product is not a chiral field.

\end{itemize}

%=================================================================
\subsection{Dimensional Reduction from $2d$ to $0d$} \label{sec:dim-red}
%=================================================================

Let us now discuss how $2d$ $(0,2)$ gauge theories are dimensionally reduced down to $0d$ $\mathcal{N}=2$ matrix models, which we express in $0d$ $\mathcal{N}=1$ language. This class of theories provides a concrete illustration of the structures that we introduced in the previous section. Upon dimensional reduction (followed by a Wick rotation), every multiplet of the $2d$ $(0,2)$ gauge theory splits into two different $0d$ $\mathcal{N}=1$ multiplets as follows:
\begin{enumerate}

\item
2d gauge $\rightarrow$ $0d$ gaugino $+$ $0d$ chiral.

\item 
2d chiral $\rightarrow$ $0d$ chiral $+$ $0d$ Fermi.

\item 
2d Fermi $\rightarrow$ a pair of $0d$ Fermi's.

\end{enumerate}

\noindent Let us denote the $2d$ gauge, chiral and Fermi multiplets 
\begin{align}
A_{\alpha}
\; (\alpha =1,\ldots, G_{\rm 2d}) \,, \quad 
\phi_I \; (I=1,\ldots,C_{\rm 2d}) \,,
\quad 
\Lambda_A \,, \overline{\Lambda}_{\bar{A}} \; (A,\bar{A}=1,\ldots,F_{\rm 2d}) \,.
\end{align}
The $0d$ field content is 
\begin{align}
\{V_\alpha \} \,, \quad \{X_i\} = \{X_I ; Y_\alpha \} 
\,, \quad 
\{\Lambda_a \} = \{ \Lambda_A ; \bar{\Lambda}^A, \bar{\Psi}_I \}
\,.
\label{dimensional_reduction_superfields}
\end{align}
Despite using a similar notation for both $2d$ and $0d$ superfields, the distinction and correspondence between them should be clear from their subindices. From \eref{dimensional_reduction_superfields}, it follows that the number of $0d$ multiplets of each type is given by
\begin{align}
G_{\rm 0d} = G_{\rm 2d} \,,
\quad 
C_{\rm 0d} = C_{\rm 2d} + G_{\rm 2d} \,,
\quad 
F_{\rm 0d} = 2F_{\rm 2d} + C_{\rm 2d} \,.
\end{align}

In the special case of $2d$ theories on D1-branes probing toric CY 4-folds, the relation $C_{\rm 2d} - G_{\rm 2d} - F_{\rm 2d} = 0$ holds \cite{Franco:2015tna}. The $0d$ matrix models obtained by dimensionally reducing such theories then satisfy
\begin{align}
F_{\rm 0d} - 3C_{\rm 0d} + 5 G_{0d} = 0 \,.
\end{align}

Returning to general $2d$ $(0,2)$ theories, we can derive the interaction terms of the $0d$ theory from those of the $2d$ parent. The $J$-terms are 
\begin{align}
\begin{array}{|c||c|c|c|}
\hline
\ \ \ \ $0d$ \ \ \ \ & \multicolumn{3}{c|}{2d} \\ \hline
\Lambda_a & \ \ \Lambda_A \ \ & \ \ \bar{\Lambda}_{\bar{A}} = \bar{\Lambda}^A \ \ & \bar{\Psi}_{I=(\alpha\beta)} 
\\
\hline
J^a & \mathcal{J}^A & \mathcal{E}^{\bar{A}} = \mathcal{E}_A & \phi_{Y_{\alpha\alpha}} \phi_{X_{\alpha\beta}} - \phi_{X_{\alpha\beta}} \Phi_{Y_{\beta\beta}} \\ \hline
\end{array}
\end{align}
In the last column, we used the standard quiver notation for bifundamental and adjoint fields.\footnote{Of course there can be more than one field for every pair of subindices. We leave this possibility implicit in order to simplify the notation.} Below we will use this notation whenever it is helpful. We also adopted a convention in which a barred superscript is the same as an unbarred subscript and vice versa. 

\begin{figure}[htbp]
	\centering
	\includegraphics[height=8cm]{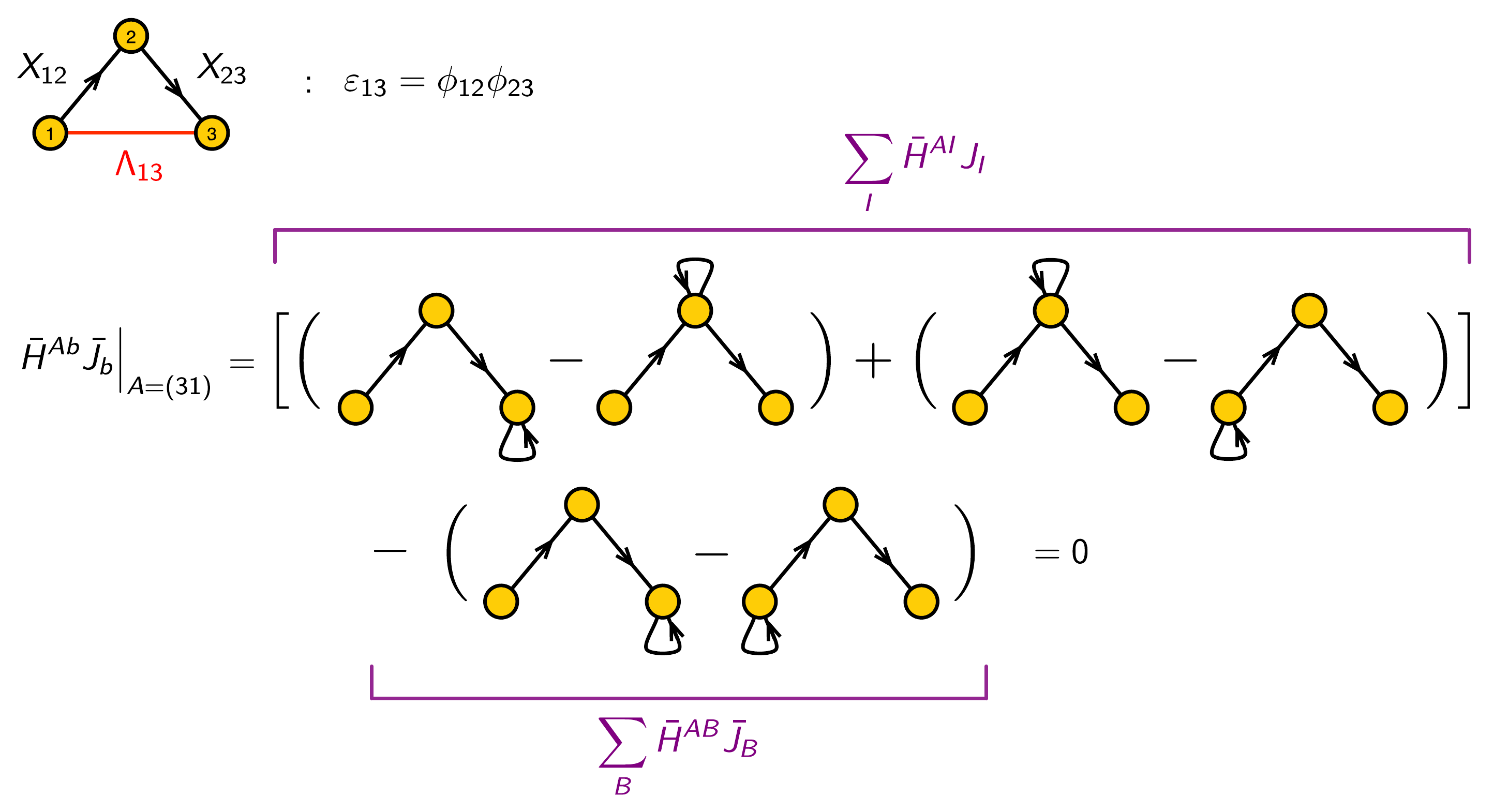}
\caption{Graphical representation of the vanishing of the $H$-constraints for a $(\Lambda_A,\bar{\Lambda}_{\bar{A}})$ pair.}
	\label{fig:H-dim-red}
\end{figure}

The $H$-terms of the $0d$ theory are 
\begin{align} 
\overline{H}^{AI} = \frac{\partial \overline{\mathcal{E}}^A }{\partial \bar{\phi}_{X_I}} \,,
\quad 
\overline{H}^{\bar{A}I} = \frac{\partial \overline{\mathcal{J}}^{\bar{A}}}{\partial \bar{\phi}_{X_I}} \,,
\quad
\overline{H}^{A \bar{B}} =  - \left. \delta^{A\bar{B}} (\bar{\phi}_{Y_{\alpha\alpha}} - \bar{\phi}_{Y_{\beta\beta}}) \right|_{A=(\alpha\beta)} \,.
\end{align}
The $H$-constraint \eqref{H-constraint} for $\bar{\Psi}_I$ is automatically satisfied due to the trace condition of the $2d$ theory
\begin{align}
\sum_A (\bar{J}_A \overline{H}^{AI} + \bar{J}_{\bar{A}}\overline{H}^{\bar{A}I}) 
=  
\sum_A 
\left(
\overline{\mathcal{J}}_A \frac{\partial \overline{\mathcal{E}}^A }{\partial \bar{\phi}_{X_I}}
 + \overline{\mathcal{E}}_{\bar{A}} \frac{\partial \overline{\mathcal{J}}^{\bar{A}}}{\partial \bar{\phi}_{X_I}}
\right)
= 
\frac{\partial}{\partial \bar{\phi}_{X_I}} 
\left( \sum_{A} \overline{\mathcal{J}}_A  \overline{\mathcal{E}}^A \right) = 0 \,.
\end{align}
The $H$-constraints for $(\Lambda_A,\bar{\Lambda}_{\bar{A}})$ schematically read as follows 
{\small
\begin{align}
\left( 
\sum_I \overline{H}^{AI} \bar{J}_I + \sum_B \overline{H}^{A\bar{B}} \bar{J}_{\bar{B}}  
\right)_{A =(\alpha\beta)}  
= 
\sum_{\bar{\phi} \in \overline{\mathcal{E}}} \frac{\partial \overline{\mathcal{E}}^{\alpha\beta}}{\partial \bar{\phi}_{X_{\gamma\delta}}} 
( \bar{\phi}_{Y_{\gamma\gamma}} \bar{\phi}_{X_{\gamma\delta}} - \bar{\phi}_{X_{\gamma\delta}} \bar{\phi}_{Y_{\delta\delta}} ) 
- (\bar{\phi}_{Y_{\alpha\alpha}} - \bar{\phi}_{Y_{\beta\beta}}) \overline{\mathcal{E}}^{\alpha\beta} \,.
\end{align}}
\fref{fig:H-dim-red} shows how the two partial sums cancel against each other.

%=================================================================
%=================================================================

%=================================================================
\subsection{Gauge/Global Symmetries and Anomalies \label{sec:anomaly}}
%=================================================================

A $0d$ theory contains no derivatives. So, we cannot gauge a symmetry by turning an ordinary derivative into a covariant derivative. It may seem unclear whether/how we can distinguish a gauge symmetry from a global symmetry.

Guided by dimensional reduction from higher dimensions, we distinguish gauge symmetries from global symmetries in three ways. First, we assign a gaugino multiplet for each factor in the gauge symmetry. There is no multiplet associated to the global symmetry. Second, we require that all observables (more on this below) be gauge invariant, but may be charged under global symmetry. Third, we require that the ``anomaly" for the gauge symmetry vanishes, but allow for non-vanishing global anomaly. By anomaly we mean any non-invariance of the integration measure under the symmetry action. 

In the presence of an anomaly, gauge or global, the integration over the symmetry orbit forces the partition function (i.e. the expectation value of the identity operator) to vanish. Since all observables are gauge invariant, a gauge anomaly implies that the theory is completely trivial. In contrast, a global anomaly can be cancelled by computing correlation functions of charged observables.

In $0d$, anomalies are linear. We will use the matching of abelian flavor anomalies, in the spirit of 't Hooft anomaly matching in higher dimensions, when we argue for the equivalence of several theories. The integration measure should carry a definite charge, i.e. anomaly, under global symmetries. If two matrix models are dual, as a necessary condition, their global anomaly should be the same.

%=================================================================
\subsection{Observables} 
%=================================================================

The superalgebra $Q^2=0$ suggests that the observables of the theory are cohomology classes of $Q$. We begin with the ring of all gauge invariant products of elementary fields and study the cohomology of $Q$. The resulting observables, 
\begin{align}
O_i \in \frac{\mbox{ker}(Q)}{\mbox{Im}(Q)} \,, 
\end{align}
still form a ring. By a slight abuse of language, we may call it the ``chiral ring". 

Suppose we enumerate all possible $O_i$'s. In principle, solving the theory completely means calculating arbitrary correlation functions among them,  
\begin{align}
\langle O_{i_1} O_{i_2} \cdots O_{i_n} \rangle 
= \int \mathcal{D}(\mbox{fields}) O_{i_1} O_{i_2} \cdots O_{i_n} e^{-S} \,.
\end{align}
Here, the classical action, the integral measure and the observables are all independently gauge invariant. As for the global symmetries, we leave the possibility that the integral measure has some abelian anomaly, which can be cancelled by the global charges of the observables. 

In summary, if we want to establish duality between two theories, we should check the following things: (1) global anomaly, (2) spectrum of observables (including their global charge and $R$-charge), (3) correlation functions. 

%=================================================================
\section{Geometric Motivation for Quadrality}
%=================================================================

\label{section_geometric_motivation}

In this section we use the geometric engineering of $0d$ $\mathcal{N}=1$ theories in terms of D(-1)-branes probing toric CY 5-folds to motivate a new duality. This duality turns out to be of order 4, so we will refer to it as {\it quadrality}. 

%=================================================================
\subsection{D-Branes Probing Calabi-Yau Manifolds}
%=================================================================

\label{section_D-brane_probes}

The $(10-2n)$-dimensional gauge theories that live on the worldvolume of D$(9-2n)$-branes probing toric CY $n$-folds have been studied in great detail for $n=3$ and $4$. For toric CYs, the connection between gauge theory and geometry is considerably simplified by T-dual brane configurations: brane tilings for CY$_3$ \cite{Franco:2005rj,Franco:2005sm} and brane brick models for CY$_4$ \cite{Franco:2015tna,Franco:2015tya,Franco:2016nwv}. It is natural to continue with this sequence as summarized in Table \ref{sequence_theories} and consider $n=5$, which corresponds to D(-1)-branes probing toric CY 5-folds. The theories on their worldvolume are $0d$ $\mathcal{N}=1$ gauged matrix models of the type discussed in section \sref{sec:super-matrix}.

%=================================================================
\begin{table}[h]
\begin{center}
\begin{tabular}{|c|c|c|}
\hline
{\bf QFT} & {\bf Original geometry} & {\bf Mirror} \\ \hline
$4d$ $\mathcal{N}=1$ & IIB D3 probing CY$_3$ & IIA D6 on 3-cycles \\ \hline
$2d$ $(0,2)$ & IIB D1 probing CY$_4$ & IIB D5 on 4-cycles \\ \hline
$0d$ $\mathcal{N}=1$ (matrix model) & IIB D(-1) probing CY$_5$ & IIB ED4 on 5-cycles \\ \hline
\end{tabular}
\end{center}
\vspace{-.3cm}\caption{Configurations of D-branes giving rise to quantum field theories in various dimensions.}
\label{sequence_theories}
\end{table}
%=================================================================

T-duality relates D(-1)-branes probing toric CY 5-folds to Type IIA configurations analogous to brane tilings and brane brick models, to which we refer as {\it brane hyperbrick models}. A brane hyperbrick model consists of an NS5-brane wrapped over a holomorphic surface $f(w,x,y,z)=0$, from which stacks of Euclidean D3-branes are suspended. The whole configuration lives on a $T^4$. Very much like their lower dimensional cousins, we expect brane hyperbrick models to provide valuable tools for connecting geometry to gauge theory. However, they are beyond the scope of this paper and we leave their development for future work.

%=================================================================
\subsection{Mirror Symmetry and D-branes at Toric Singularities}
%=================================================================

\label{section_mirror_general}

In order to set up the stage for the discussion in the coming section, we present a lightning review of the mirror approach to D-branes probing toric CY singularities. We refer the reader to \cite{Feng:2005gw,Futaki:2014mpa,Franco:2016qxh} for details.

A toric CY$_n$ $\mathcal{M}$ is specified by its toric diagram $V$, which is a convex set of points in $\mathbb{Z}^{n-1}$. Its mirror geometry \cite{Hori:2000kt,Hori:2000ck} is an $n$-fold $\mathcal{W}$ given by a double fibration over the complex $W$ plane
\beq
\begin{array}{rl}
W = & P(x_1,\ldots, x_{n-1}) \\[.1cm]
W = & uv
\end{array}
\label{double_fibration}
\eeq 
with $u,v \in \mathbb {C}$ and $x_\mu\in \mathbb{C^*}$, $\mu=1,\ldots,n-1$. $P(x_1,\ldots, x_{n-1})$ is the Newton polynomial
\beq
P(x_1,\ldots, x_{n-1})=\sum_{\vec{v} \in V} c_{\vec{v}} \, x_1^{v_1} \ldots x_{n-1}^{v_{n-1}} , 
\eeq
where the $c_{\vec{v}}$ are complex coefficients and we sum over points $\vec{v}$ in the toric diagram. By rescaling the $x_\mu$ variables, it is possible to set $n$ of the coefficients to 1. 

The critical points of $P$ are $(x_1^*,\ldots,x_{n-1}^*)$ such that
\beq
\left. {\partial \over \partial x_\mu} P(x_1,\ldots, x_{n-1})\right |_{(x_1^*,\ldots,x_{n-1}^*)}=0 \ \ \ \ \forall \, \mu \, .
\eeq
They correspond to critical values $W^*=P(x_1^*,\ldots,x_{n-1}^*)$. The number of critical points is equal to the number of gauge nodes in the field theory \cite{Feng:2005gw}.

The two fibers are a holomorphic $(n-2)$ complex dimensional surface $\Sigma_W$ coming from $P(x_1,\ldots, x_{n-1})$ and a $\mathbb{C}^*$ fibration associated with the $uv$ piece. The resulting $S^{n-2}\times S^1$ fibration over a straight {\it vanishing path} that stretches between $W=0$ and $W=W^*$ hence gives rise to an $S^n$.\footnote{Vanishing paths can be curved. We refer the reader to \cite{Franco:2016qxh} for a discussion of this possibility. This subtlety does not affect our conclusions.}

The $S^1$ fiber vanishes at $W=0$. The structure of the gauge theory is determined by how the surviving $S^{n-2}$'s intersect on the vanishing locus $W^{-1}(0):P(x_\mu)=W=0$. The geometry of the $S^{n-2}$'s can be efficiently described using {\it tomography}, which was introduced in \cite{Futaki:2014mpa} and further developed in \cite{Franco:2016qxh}. The $x_\mu$-tomography is the projection of the $S^{n-2}$ spheres at $W=0$ onto the $x_\mu$-plane.  In summary, we can obtain a detailed description of the configuration of $S^n$ in the mirror geometry by combining the configuration of vanishing paths on the $W$-plane with the $x_\mu$-tomographies, $\mu=1,\ldots,n$.

The discussion in the coming section will build on ideas introduced in \cite{Franco:2016qxh}. It will just use basic properties of vanishing paths on the $W$-plane, which is a universal ingredient for any CY dimension.

%=================================================================
\subsection{Quadrality from Mirror Symmetry}
%=================================================================

\begin{figure}[ht]
	\centering
	\includegraphics[width=13cm]{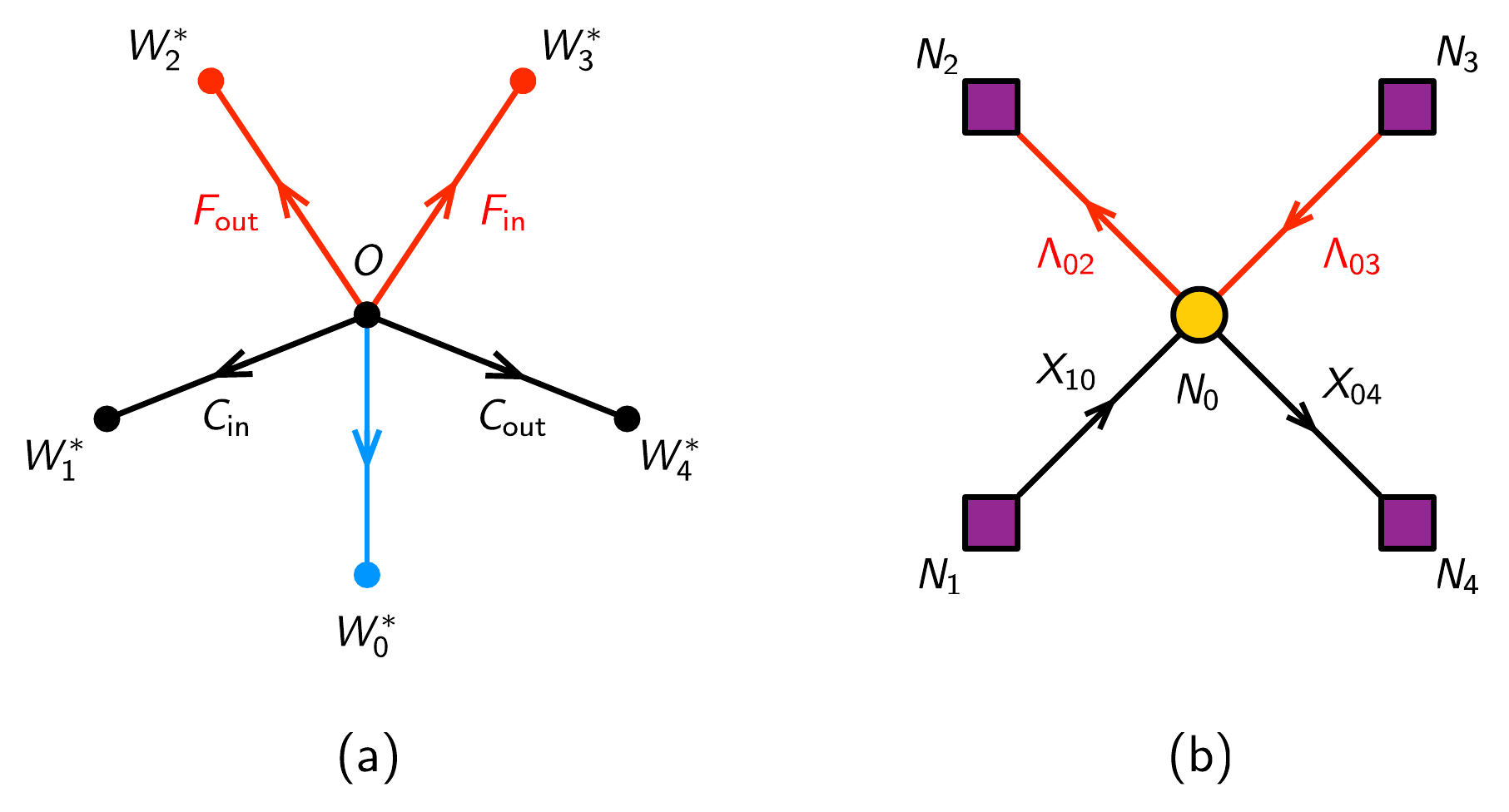}
\caption{a) Configuration of vanishing paths on the $W$-plane for a CY$_5$ for a reference cycle associated to $W_0^*$. We indicate the type of flavor contributed by each cycle: chiral ($C_{in}$ and $C_{out}$) or Fermi ($F_{in}$ and $F_{out}$). b) Corresponding quiver diagram. We refer to this theory as $T$.}
	\label{mirror_quadrality}
\end{figure}

Following the discussion in the previous section, mirror symmetry maps D(-1)-branes probing a toric CY$_5$ to a collection of Euclidean D4-branes wrapping 5-cycles. Let us consider some elementary features of this mirror configuration. Given a vanishing cycle $C_0$, the branes wrapping the cycles that intersect with it can be regarded as flavor branes. These flavor cycles can be classified into four groups, depending on the type of matter fields they contribute to $C_0$: fundamental and antifundamental chirals and fundamental and antifundamental Fermis. Following the quiver representation of these fields, we refer to fundamental and antifundamental representations as {\it in} and {\it out}, respectively. Extending the ideas introduced in \cite{Franco:2016qxh}, we expect that for any $C_0$ the vanishing paths that are associated to the four types of flavors are organized cyclically on the $W$-plane as shown in \fref{mirror_quadrality}. This conjecture is a natural generalization of what occurs for CY 3- and 4-folds and is also based on the expected symmetry between quadrality and inverse quadrality.

%=================================================================
%=================================================================

\begin{figure}[ht]
	\centering
	\includegraphics[width=15cm]{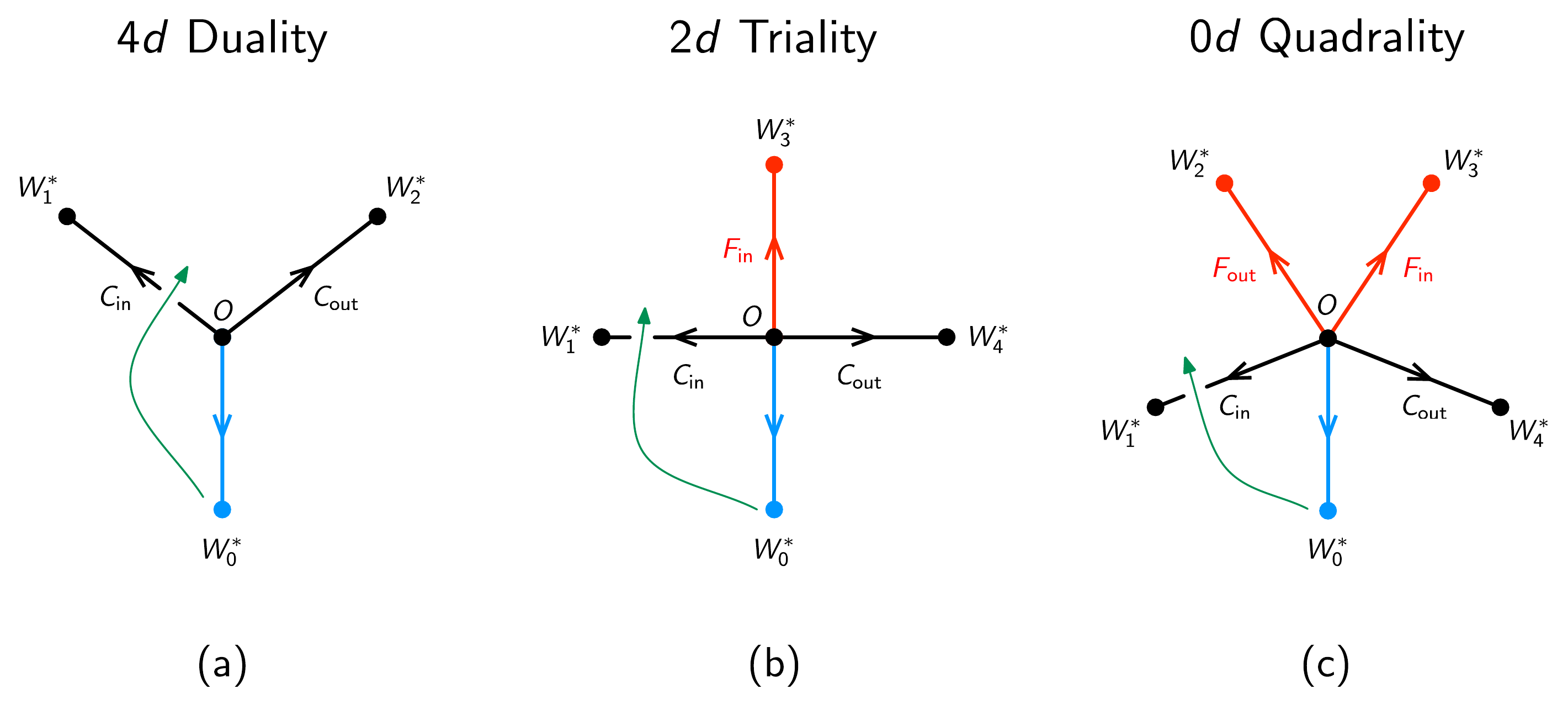}
\caption{Unified description of QFT equivalences in various dimensions. Mirror symmetry unifies them in terms of the same type of transformation.}
	\label{mirror_dualities}
\end{figure}

An important lesson from \cite{Cachazo:2001sg,Franco:2016qxh} is that mirror symmetry not only provides a geometric unification of dualities for a fixed dimension, but also unifies seemingly different QFT equivalences across dimensions. \fref{mirror_dualities}.a and b show that the mirror realizations of $4d$ Seiberg duality and $2d$ triality are basically identical. This observation naturally leads to conjecture that the transformation in \fref{mirror_dualities}.c gives rise to an equivalence between $\mathcal{N}=1$ gauged matrix models that we call quadrality.

%=================================================================
%=================================================================

Let us denote $Q_i$ the D-brane charge vector for the cycle $C_i$. The intersection matrix $\langle C_i,C_j\rangle=\langle Q_i,Q_j\rangle$ of a CY$_5$ is antisymmetric. The sign of $\langle Q_i,Q_j\rangle$ determines the orientation of the bifundamental(s) connecting nodes $i$ and $j$. The distinction between chiral and Fermi fields requires additional information regarding the intersection. Without loss of generality, let us assume that the only non-vanishing intersections of cycles are
\beq
\begin{array}{ccccccc}
\langle Q_1,Q_0\rangle & = & 1 & \ \ \ \ \ \ \ \ & \langle Q_2,Q_0\rangle & = & -1 \\
\langle Q_3,Q_0\rangle & = & 1 & \ \ \ \ \ \ \ \ & \langle Q_4,Q_0\rangle & = & -1 
\end{array}
\eeq
In our discussion, any multiplicities of different types of flavors are absorbed into the ranks of the corresponding flavor nodes. Introducing more general intersection numbers $\langle Q_i,Q_0 \rangle\neq \pm 1$ or splitting the flavor nodes into collections of multiple nodes with different brane charges is straightforward. Non-vanishing intersections between the flavor nodes in the initial theory, corresponding to additional matter fields, can also be incorporated without affecting our analysis.

Quadrality on node 0 corresponds, as shown in \fref{mirror_dualities}, to shrinking the cycle $C_0$ to zero size and reemerging on the $W$-plane on the wedge between $C_1$ and $C_2$ with a reversed orientation. The brane charges transform as follows:
\beq
\begin{array}{ccl}
Q_0' & = & - Q_0 \\[.15cm]
Q_{1}'  & = & Q_1 +  \langle Q_1,Q_0\rangle Q_0 = Q_{1} + Q_0 \\[.15cm]
Q_{2}' & = & Q_2 \\[.15cm]
Q_{3}' & = & Q_3 \\[.15cm]
Q_4' & = & Q_4 
\end{array}
\label{change charges}
\eeq
Inverse quadrality is obtained by moving $C_0$ in the opposite direction to the wedge between $C_3$ or $C_4$, or by applying quadrality three times.

Remarkably, even without a more detailed analysis of the mirror, we can draw important conclusions about what this transformation implies for quadrality.

%=================================================================
\paragraph{Rank of the Gauge Group.}
%=================================================================

The transformation of the rank of the gauge group follows from conservation of the total brane charge. Initially, we have
\beq
Q_T = \sum_{i=0}^4 N_i \, Q_i \, .
\eeq
Since the ranks of the flavor nodes remain constant, we have
\beq
\begin{array}{ccl}
Q_T' & = & \sum_{i=0}^4 N_i' \, Q_i'  \\[.15cm]
& = & -N_0' \, Q_0 + N_1 \, (Q_1+Q_0)+N_2\, Q_2+N_3\, Q_3+N_4 \, Q_4 \\[.15cm]
& = & Q_T + \left[(N_1-N_0)-N_0'\right] Q_0 \, .
\end{array}
\eeq
Conservation of brane charge $Q_T=Q_T'$ implies
\beq
N_0'=N_1-N_0 \, .
\eeq

%=================================================================
\paragraph{Dual Flavors.}
%=================================================================

\begin{figure}[ht]
	\centering
	\includegraphics[width=10cm]{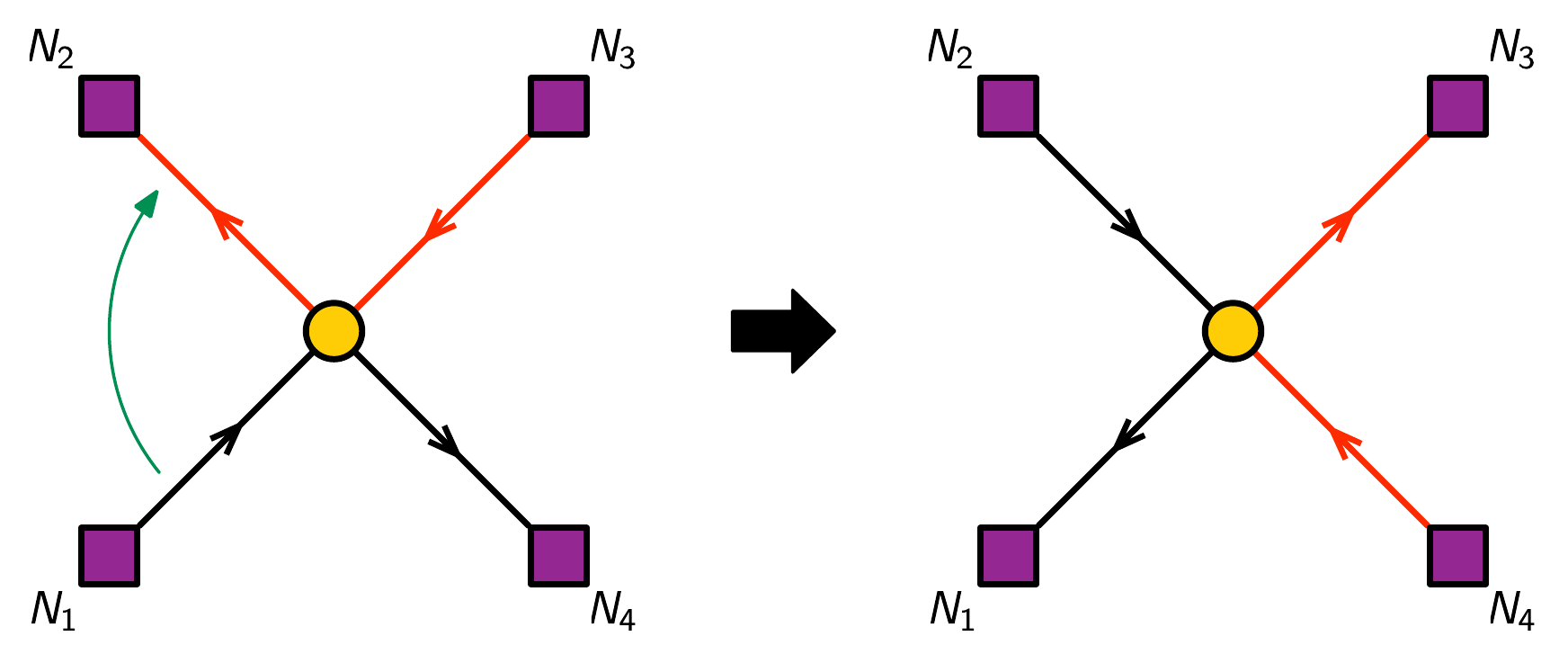}
\caption{The transformation of flavors under quadrality is a rotation of the corresponding arrows by $\pi/2$, while keeping the flavor nodes fixed.}
	\label{rotation_flavors}
\end{figure}

The flavor vanishing paths divide the $W$-plane into four wedges. Quadrality corresponds to moving the cycle associated to the dualized gauge group to a neighboring wedge.  This implies that, in the quiver, the transformation of flavors is a $\pi/2$ rotation of the corresponding arrows, while keeping the flavor nodes fixed. This is shown in \fref{rotation_flavors}

%=================================================================
%=================================================================

This is yet another manifestation of the geometric unification of QFT equivalences in different dimensions The transformations of flavors in Seiberg duality and triality are also given by $2\pi/k$ rotations, with $k=2,3$, respectively \cite{Cachazo:2001sg,Franco:2016qxh}.

%=================================================================
\paragraph{Mesons.}
%=================================================================

The modification of $Q_1$ in \eref{change charges} leads to non-vanishing intersections between node 1 and nodes 2, 3 and 4 in the dual. These new intersections give rise to mesons.  Furthermore,
\beq
\begin{array}{ccccc}
\langle Q_1',Q_2 \rangle & = & \langle Q_0,Q_2\rangle & \ \to \ \ & \mu_{12} \, \mbox{(Fermi)} \\[.15cm]
\langle Q_1',Q_3 \rangle & = & \langle Q_0,Q_3\rangle & \ \to \ \ & \mu_{31} \, \mbox{(Fermi)} \\[.15cm]
\langle Q_1',Q_4 \rangle & = & \langle Q_0,Q_4\rangle & \ \to \ \ & M_{14} \, \mbox{(chiral)} 
\end{array}
\label{new_mesons_mirror}
\eeq
The chirality and type of field for each meson is thus determined by those of the original flavors connecting node 0 to the global symmetry nodes. We conclude the dual contains the following mesons: $\mu_{12}$ (Fermi), $\mu_{31}$ (Fermi) and $M_{14}$ (chiral). This expectation will be confirmed by a field theory analysis in section \sref{section_quadrality_dual}.

%=================================================================
\paragraph{The Quiver.}
%=================================================================

\begin{figure}[ht]
	\centering
	\includegraphics[width=5.5cm]{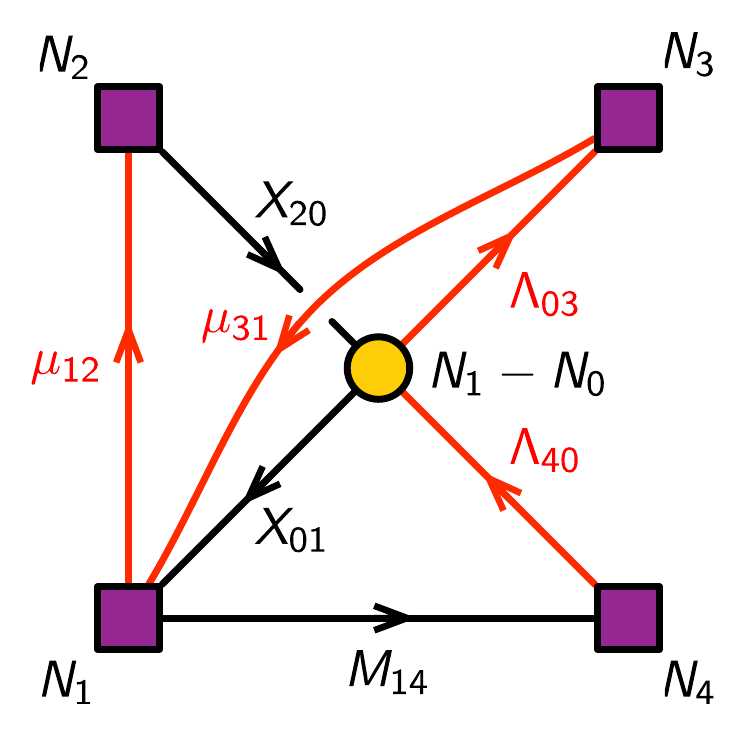}
\caption{Quiver diagram for $T'$, the quadrality dual of the theory in \fref{mirror_quadrality}.a.}
	\label{quiver_T'}
\end{figure}

Summarizing our previous analysis, we conclude that the quiver diagram for the quadrality dual of theory $T$, which we call $T'$, is the one shown in \fref{quiver_T'}. Throughout the paper, we will often adopt a notation in which $X$ and $\Lambda$ indicate chiral and Fermi flavors, from the perspective of the dualized gauge group, respectively. Similarly, $M$ and $\mu$ correspond to chiral and Fermi mesons, i.e. gauge singlets.

%=================================================================
%=================================================================

%=================================================================
\paragraph{Periodicity.}
%=================================================================

As we mentioned, it is possible to argue that quadrality is an order 4 duality based on the fact that the flavor vanishing paths divide the $W$-plane into four wedges. Here we provide a more explicit proof of the periodicity. For all $i,j=0,\ldots,4$, we have
\beq
\begin{array}{ccl}
\langle Q_i'''',Q_j''''\rangle & = & \langle Q_i +  \langle Q_i,Q_0\rangle Q_0 , Q_j +  \langle Q_j,Q_0\rangle Q_0 \rangle \\[.15cm]
& = & \langle Q_i,Q_j\rangle + \langle Q_i,Q_0\rangle \langle Q_0,Q_j\rangle +  \langle Q_j,Q_0\rangle  \langle Q_i,Q_0\rangle +  \langle Q_i,Q_0\rangle  \langle Q_0,Q_0\rangle \\[.15cm] 
& = & \langle Q_i,Q_j\rangle 
\end{array}
\eeq
where in the last line we used the antisymmetry of the intersection pairing.

%=================================================================
\subsection{Local Geometry for Order $n$ Dualities} 
%=================================================================

\label{section_local_geometry_order_n_dualities}

The minimal local geometry that captures the order $(n-1)$ duality of a $(10-2n)$-dimensional theory (i.e. $4d$ duality, $2d$ triality and $0d$ quadrality) is local $\mathbb{CP}^n$. This is the $\mathbb{C}^n/\mathbb{Z}_n$ orbifold with action $(1,\ldots,1)$ on the different complex planes. Local $\mathbb{CP}^n$ gives rise to $n$ critical points and hence to $n$ independent $n$-cycles. This number is precisely what is required for studying a generic field content in every dimension. One of the cycles accounts for the gauge group and the remaining $n-1$ corresponds to nodes (which are also gauged) for all possible types of flavors: two in $4d$ (fundamental and antifundamental chirals), three in $2d$ (fundamental and antifundamental chirals, and Fermi) and four in $0d$ (fundamental and antifundamental chirals, and fundamental and antifundamental Fermis).

\begin{figure}[ht]
	\centering
	\includegraphics[width=12cm]{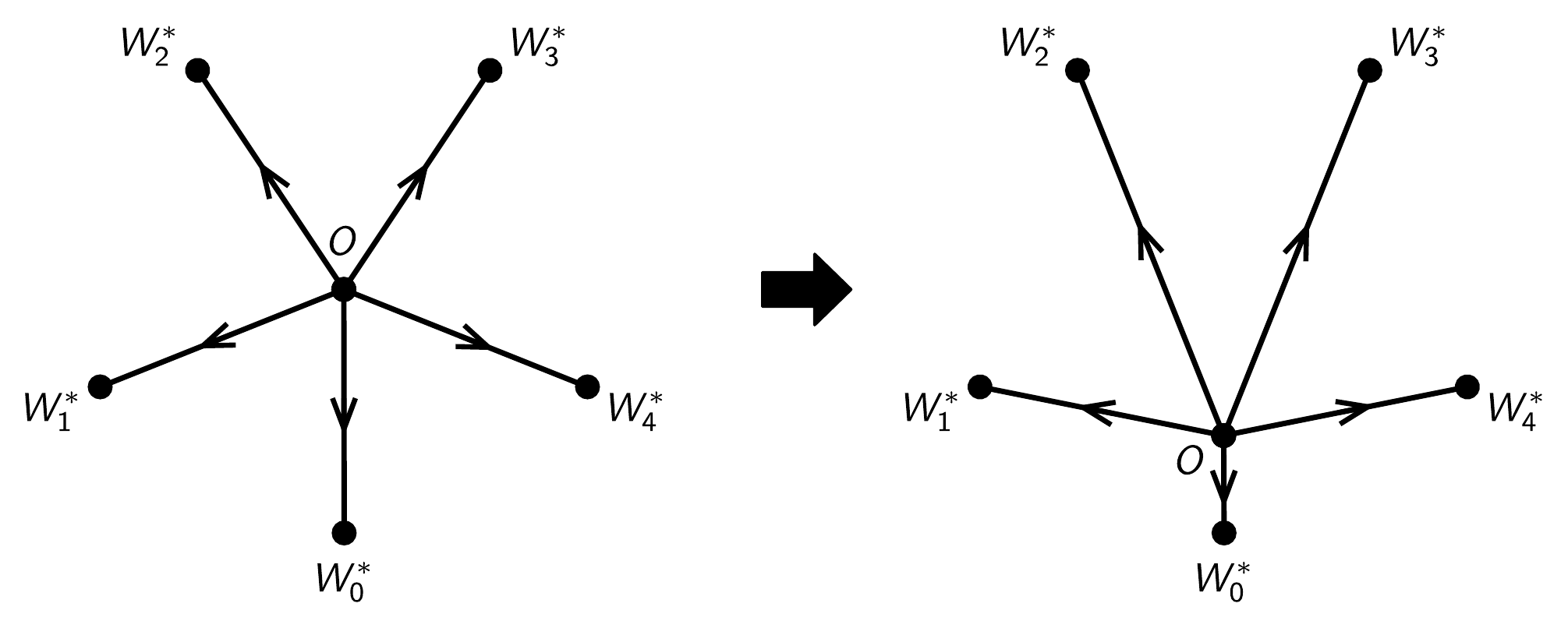}
\caption{By sending $\alpha \to W_i^*$ we can make the corresponding $n$-cycle, here $C_0$, shrink to zero volume.}
	\label{strong_coupling_CP4}
\end{figure}

The Newton polynomial coming for the toric diagram for local $\mathbb{CP}^n$ contains, in general, $n+1$ terms. It is possible to rescale $n$ of its coefficients to $1$, leading to
\beq
P(x_1,\ldots,x_n)=x_1+\ldots+x_n+{1\over x_1 \ldots x_n} + \alpha \, ,
\label{Newton_polynomial_CPn}
\eeq
with $\alpha \in \mathbb{C}$. The parameter $\alpha$ does not affect the critical points $\phi_i^*$. However, it modifies the corresponding critical values $W_i^*$ by an overall shift. Equivalently, we can think that the critical values $W_i^*$ are fixed and that the origin of the $W$-plane is shifted by $\alpha$. Thus, this geometry has enough freedom for studying the limit in which any of the $n$ gauge groups go to infinite coupling. We can attain this by sending $\alpha \to W_i^*$ as shown in \fref{strong_coupling_CP4} for $\mathbb{CP}^5$, which makes the volume of the corresponding cycle vanish.

%=================================================================
%=================================================================

%=================================================================
\section{Quadrality}
%=================================================================

\label{section_quadrality}

\begin{figure}[ht]
	\centering
	\includegraphics[width=13cm]{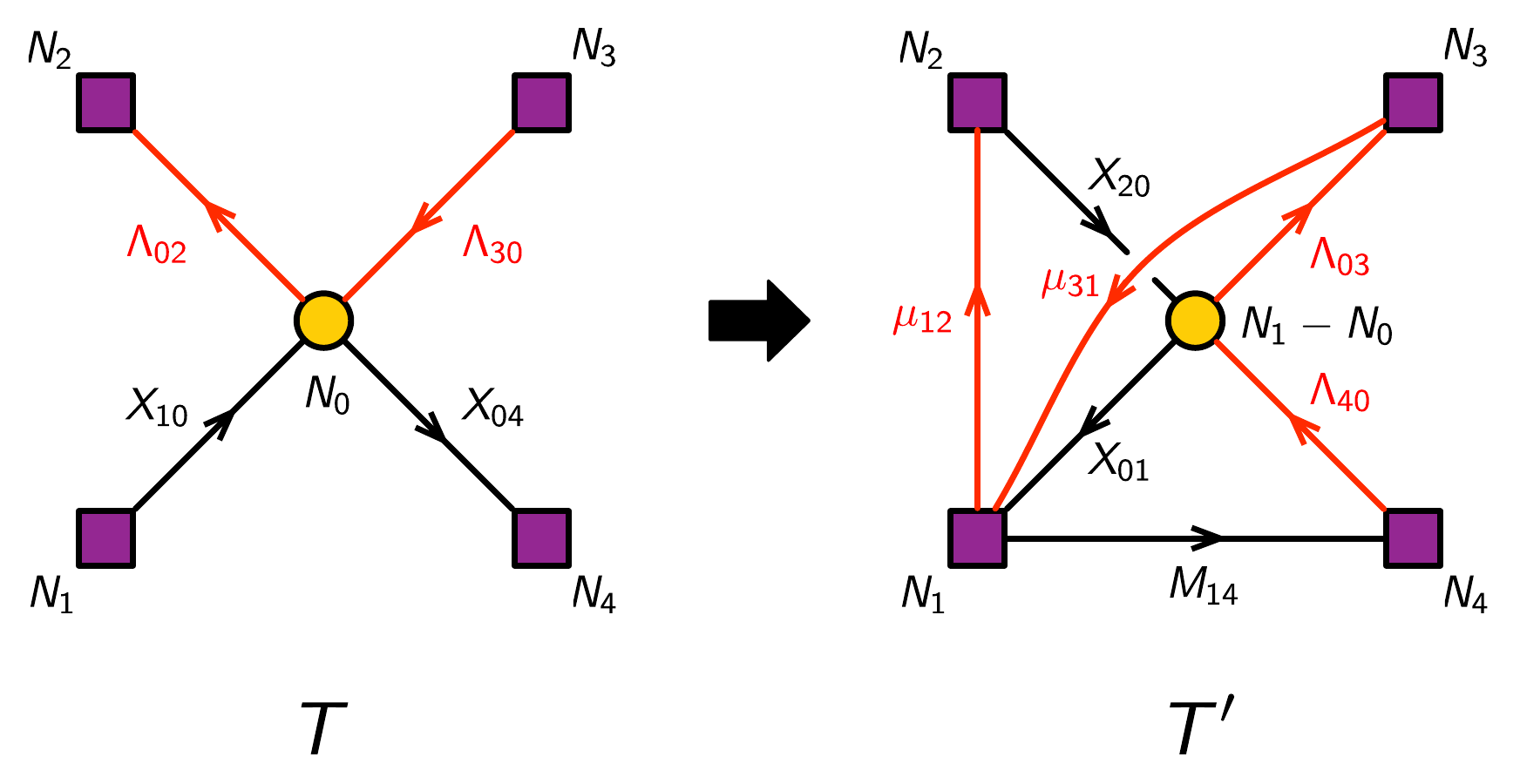}
\caption{Quiver diagrams for the initial theory $T$ and its quadrality dual $T'$.}
	\label{quivers_T_T'}
\end{figure}

In the previous section we motivated quadrality using mirror symmetry for configurations of D(-1)-branes probing toric CY 5-folds. It is however natural to conjecture that, as it occurs for Seiberg duality and triality, that quadrality applies to arbitrary $\mathcal{N}=1$ gauged matrix models. The elementary quadrality transformation, which we now explain in more detail, can be phrased in terms of the simple SQCD-like theory 
$T$ that we introduced earlier. For quick reference, \fref{quivers_T_T'} reproduces the quiver diagrams for $T$ and its quadrality dual.

%=================================================================
%=================================================================

Let us discuss $T$ first. The four flavor nodes and the gauge node in the quiver correspond to $U(N_i)=SU(N_i)\times U(1)^{(i)}$ groups. All matter fields transform in bifundamental representations, so the global diagonal combination of all of them, $\sum_{i=0}^4Q_i$ decouples.\footnote{From now on $Q_i$ refers to the charge under $U(1)^{(i)}$. It should not be confused with the D-brane charges discussed earlier.} Without loss of generality, we can identify the global symmetry with $SU(N_1)\times SU(N_2)\times SU(N_3)\times SU(N_4)\times U(1)^{(1)}\times U(1)^{(2)}\times U(1)^{(3)}$.\footnote{$U(1)^{(4)}$ is not independent, and the corresponding charge is simply $Q_4=-(Q_0+Q_1+Q_2+Q_3)$.} It is straightforward to read the transformation properties of the matter fields under the global symmetry group from the quiver.

Cancellation of the abelian gauge anomaly constrains the ranks of the flavor nodes to satisfy
\beq
N_1-N_2+N_3-N_4=0 \, .
\label{abelian_gauge_anomaly}
\eeq
If more nodes are gauged, as in the D-branes examples considered in section \sref{section_D-brane_examples}, there will be additional matter fields that ensure the cancellation of all gauge anomalies.

%=================================================================
\subsection{The Quadrality Dual}
%=================================================================

\label{section_quadrality_dual}

We propose the quadrality dual $T'$ is given by the quiver in \fref{quivers_T_T'}, together with some $J$- and $H$-terms that we discuss below. Since $N_1+N_3-N_2-N_4=0$, the abelian gauge anomaly still vanishes. The global symmetry of the dual is $SU(N_1)\times SU(N_2)\times SU(N_3)\times SU(N_4)\times U(1)^{(1)}\times U(1)^{(2)}\times U(1)^{(3)}$, in agreement with the original theory.\footnote{When determining the global symmetry of a theory, it is necessary to take into account its $J$- and $H$-terms, which can in principle break the naive symmetries preserved by the quiver. We discuss these interactions below.}

Let us explain the arguments that lead to this proposal.

%=================================================================
\paragraph{Dual Gauge Group.}
%=================================================================

The dual theory has a $U(N_0')$ gauge symmetry, with
\beq
N_0'=N_1-N_0 .
\eeq
This result was derived in section \sref{section_geometric_motivation} in the case of theories with a D-brane realization from conservation of the total brane charge. We postulate it holds in general.

%=================================================================
\paragraph{Dual Flavors.}
%=================================================================

For theories arising on D-branes, we used mirror symmetry to derive the transformation of flavors summarized in \fref{rotation_flavors}. Once again, we propose this transformation applies to general theories.

%=================================================================
\paragraph{Mesons.}
%=================================================================

There are three types of mesons in $T'$. They can be expressed as composites of the fields in the original theory. In all cases they must contain $X_{10}$ which, for D-branes, is the chiral field charged under the flavor node whose brane charge changes under quadrality. The mesons are given by
\beq
\begin{array}{ccccc}
M_{14} & = & X_{10} \, X_{04} & \ \ \ & \mbox{(chiral)} \\
\mu_{12} & = & X_{10} \, \Lambda_{02} & \ \ \ & \mbox{(Fermi)} \\
\mu_{31} & = & \Lambda_{30} \, \overline{X}_{10} & \ \ \ & \mbox{(Fermi)} 
\end{array}
\eeq
The types of fields obtained by taking these products nicely coincide with the ones established using the brane intersections in \eref{new_mesons_mirror}. Notice that in order to form the gauge invariant meson $\mu_{31}$, it is necessary to conjugate $X_{10}$. This is a novel feature of $0d$, which does not arise in $4d$ or $2d$.

%=================================================================
\paragraph{Dual Flavors-Meson Couplings.}
%=================================================================

As it occurs in Seiberg duality and triality, there are new interaction terms coupling the mesons to the dual flavors. These terms are the most general ones allowed by the gauge and global symmetries.\footnote{This principle also holds for more complicated theories. We should always include all interactions allowed by the gauge and global symmetries.} In this case, the couplings are
\beq
\begin{array}{rclcc}
& & & & \mbox{{\bf Quiver loop}} \\[.15cm]
J_{\mu_{12}} & = & X_{20} X_{01} & \ \ \ \ \to \ \ \ \ & (\mu_{12} X_{20} X_{01}) \\[.16cm]
J_{\Lambda_{40}} & = & X_{01} M_{14} & \ \ \ \ \to \ \ \ \ & (\Lambda_{40} X_{01} M_{14}) \\[.08cm]
\overline{H}^{\Lambda_{03},\mu_{31}} & = & \overline{X}_{01} & \ \ \ \ \to \ \ \ \ & (\Lambda_{03}\, \mu_{31} \overline{X}_{01}) 
\end{array}
\eeq
To simplify visualization of the interactions, on the right column we give the corresponding loops in the quiver. Below we will explain how they are crucial for consistency of quadrality.

Interestingly, the appearance of a novel type of meson $\mu_{31}= \Lambda_{30} \, \overline{X}_{10}$ is correlated with the existence of a new type of interactions in $0d$, the $H$-terms.

%=================================================================
\subsection{The Meaning of Quadrality in $0d$}
%=================================================================

In this section we would like to examine what we can mean by duality in the context of a $0d$ QFT.  Typically when we have a Seiberg-like duality we start with two distinct theories in the UV, and only in the IR the two theories will become identical, as they flow to the same conformal fixed point.  In such situations, some aspects, such as the chiral ring, are invariants of the flow and can be studied on either side of the duality even before the flow.  Checking this match has been one of the key evidences for Seiberg-like dualities.  But duality extends beyond the chiral sector and is expected that at the IR fixed point, correlations function of any collection of fields, whether chiral or not, match on both sides, with a suitable dictionary of how operators from one side map to the other.

In the case at hand, a robust check of our proposed quadrality is to verify that the chiral rings match for each dual version.  However this is not enough to claim equivalence of two theories, as there are non-chiral operators in the theory.  So one would like to have the analog notion of ``IR" in such theories, so that one could say that the IR of all sides agree for all operators.  However, there is difficulty defining the notion of IR fixed point in the present context, because the dimension of space-time is 0 and so we have no such notion.  Instead we propose the following alternative. As a supersymmetric theory flows to the IR in $d>0$ cases, chiral fields do not get renormalized (which we will loosely call ``F-terms") but non-chiral fields do get renormalized (the ``D-terms").  So it is natural to define the notion of IR in $d=0$ theories by saying that there is a deformation of the ``D-terms" in the Lagrangian (i.e. $Q$-trivial additions to the Lagrangian) which leads the theory to have the expected superconformal symmetry.
So we need to address what is the superconformal group in $0d$.

It is natural to expect that the IR superconformal field theory (SCFT) has a $U(1)$ R-symmetry. Then, the superconformal group in $0d$ should have a bosonic conformal symmetry given by $SO(1,1)\times U(1)_R$.  Let us call the generators of this group $\Delta$ and $R$.  Moreover we expect, as in higher dimensions, that the number of supersymmetries gets doubled at the conformal point.  Since we started with one nilpotent supercharge, let us call it $Q^+$, we should obtain another one, $Q^-$. We expect the following superconformal algebra:
\beq
\begin{array}{ccccc}
 Q^{+2}=Q^{-2}=0 \, , & \ \ \ \ \ \ & [\Delta,R]=0 \, , & \ \ \ \ \ & \{Q^+,Q^-\}=2\Delta -  R \,, \\[.15cm] 
 [\Delta, Q^{\pm} ]= \pm \frac{1}{2} Q^{\pm} \,,  & & [R, Q^{\pm} ]= \pm  Q^{\pm}  \,. & & 
\end{array}
\eeq
Note that this symmetry algebra is at the level of fields in the theory and not the Hilbert space, because this theory has no time dimensions.  So we conjecture that there is a distinguished fixed point where the above algebra is a symmetry of field space and correlation functions.\footnote{It is possible (and probably likely) that the symmetry is only a symmetry of the correlation functions and not the Lagrangian itself, because the path-integral measure may not be invariant.}  Whether such a point is unique is not clear and requires further study.  In this paper we check the quadrality only by checking the chiral rings match on all sides (as is the case with duality checks in higher dimension).  This low dimension case allows us to possibly be able to check a more detailed statement by including all operators! It would be interesting to pursue this direction and see if one can precisely fix the Lagrangian at the conformal fixed point of these theories and prove the quadrality symmetry for correlation functions of all operators.

%=================================================================
\section{Checks} 
%=================================================================

\label{section_checks}

In this section we collect additional checks of the quadrality proposal.

%=================================================================
\subsection{Abelian Flavor Anomalies} 
%=================================================================

Matching of abelian flavor anomalies provides a non-trivial check of the proposal. As mentioned earlier, the non-abelian flavor anomalies trivially vanish. The following table summarizes the anomalies.

%=================================================================
\begin{center}
\begin{tabular}{|c|r|l|}
\hline
& T \ \ \ \ & \quad \quad \quad \quad \quad \quad \quad \quad \quad \quad \quad \ T' \\ \hline \hline 
\ \ \ $U(1)^{(1)}$ \ \ \ & \ \ \ $N_0$ \ \ \ & \ \ \ $N_2+N_4-N_3-N_0' = N_2+N_4-N_1-N_3+N_0 = N_0$ \ \ \ \\ \hline
$U(1)^{(2)}$ & \ $-N_0$ \ \ \ & \ \ \ $N_0'-N_1 = N_1-N_0-N_1 = - N_0$ \\ \hline
$U(1)^{(3)}$ & $N_0$ \ \ \  & \ \ \ $N_1-N_0' = N_1-N_1 +N_0 = N_0$ \\ \hline
\end{tabular}
\end{center}
%=================================================================

\noindent The $U(1)^{(4)}$ anomaly is $-N_0$. However, as explained above, it is not really independent and can be determined in terms of the other abelian anomalies.

Interestingly, the matching of the $U(1)^{(1)}$ and $U(1)^{(2)}$ anomalies between the two theories tests the existence of the non-conventional $\mu_{31}$ meson. However, it does not establish whether this field is a chiral or a Fermi. If we did not have independent derivations of the transformations of the gauge group rank based on conservation of brane charge, matching of abelian flavor anomalies would simultaneously test the combination of the rank, flavor and mesons rules.

%=================================================================
\subsection{Periodicity} 
%=================================================================

\label{section_periodicity}

\begin{figure}[htbp]
	\centering
	\includegraphics[width=9.5cm]{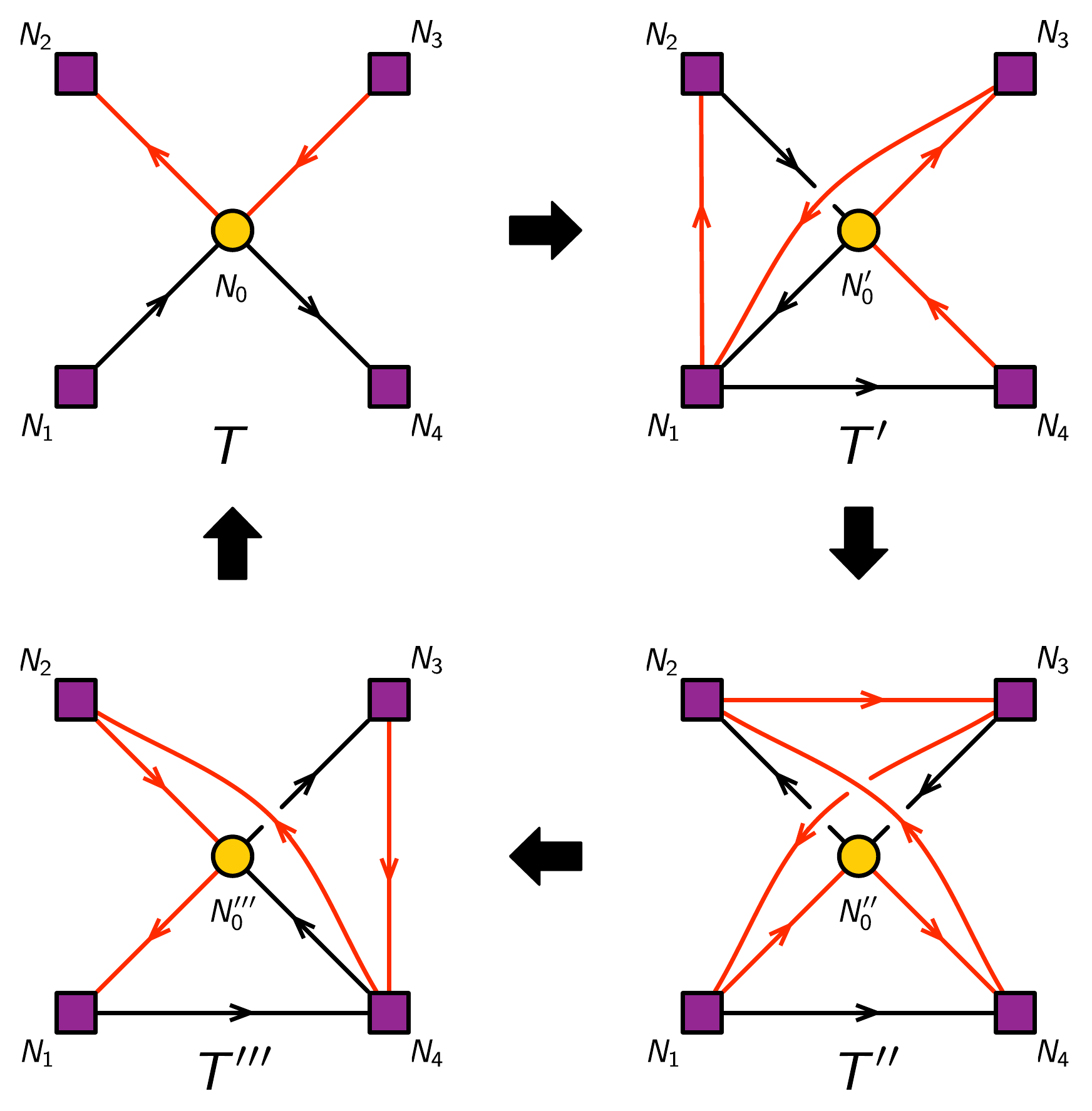}
\caption{A sequence of four quadrality transformations on the same gauge node returns to the original theory.}
	\label{quadrality_periodicity}
\end{figure}

As explained in section \sref{section_geometric_motivation}, mirror symmetry implies that after four consecutive quadrality transformations we should return to the original theory. This sequence is shown in \fref{quadrality_periodicity}. At various steps we have integrated out chiral-Fermi and Fermi-Fermi pairs, due to mass terms of the form shown in \fref{pair-cancel}. For this to be possible, the detailed form of the $J$- and $H$-terms coupling mesons to dual flavors is crucial. The periodicity of the sequence of quadralities is hence a non-trivial check of these couplings.

%======================================================================
%======================================================================

The rank of the gauge group evolves as follows
\beq
\begin{array}{ccl}
N_0' & = & N_1 - N_0 \\
N_0'' & = & N_2-N_1+N_0 \\
N_0''' & = & N_3-N_2+N_1-N_0 \\
N_0'''' & = & N_4 -N_3+N_2-N_1+N_0 = N_0
\end{array}
\label{gauge_group_rank_quadrality}
\eeq
We see that the gauge anomaly constraint on the ranks of the flavors \eref{abelian_gauge_anomaly} is crucial for returning to the original rank of the gauge group.

%=================================================================
\subsection{Deformations} 
%=================================================================

Following Seiberg's seminal work on $4d$ $\mathcal{N}=1$ duality \cite{Seiberg:1994pq}, deformations have become a standard tool for testing a wide range of equivalences between field theories. Here we will show that similar arguments can be applied to quadrality. When testing the effect of a deformation in Seiberg duality, one only needs to consider the magnetic dual. Remarkably, for quadrality we can, and actually must, study the effect of any deformation on the full quadrality sequence. This is a general feature of order $n$ dualities. The effect of deformations on the entire collection of dual theories provides an ($n$-1)-fold increase in the number of constraints and consistency checks. $\mathcal{N}=1$ matrix models have a rich set of possible deformations with which to test quadrality. In particular, we can introduce $X\Lambda$ and $\Lambda\Lambda$ mass terms, which correspond to $J$- and $H$-terms respectively.

%=================================================================
\paragraph{The Original Sequence.}
%=================================================================

The original quadrality sequence was discussed in section \sref{section_periodicity}. It consists of four theories $T$, $T'$, $T''$ and $T'''$, corresponding to the four quivers in \fref{quadrality_periodicity}. Since the global symmetry is preserved, the ranks of the flavor nodes remain equal to $(N_1,N_2,N_3,N_4)$ in all theories. The rank of the gauge group evolves according to \eref{gauge_group_rank_quadrality}.

%=================================================================
\paragraph{Possible Deformations and Deformed Sequences.}
%=================================================================

Let us consider the original theory $T$, which is shown in \fref{quivers_T_T'}. All bifundamental flavors are, generically, rectangular matrices. We can use global and gauge symmetries to simplify them, such that all entries are zero except for, at most, those in $N_0 \times N_0$ diagonal submatrices. There are three possible mass deformations of this theory:

\newpage

%=================================================================
\noindent{\bf a) \underline{$X_{10}\Lambda_{02}$ mass}} 
%=================================================================

\smallskip

$T$ can be deformed by introducing an $X_{10}\Lambda_{02}$ mass term, i.e. 
\beq
J_{\Lambda_{02}}= m \, \phi_{10} \, .
\label{deformation_a}
\eeq
 For simplicity, here and in the deformations that we discuss below, we assume the rank of the mass matrix is 1. It is straightforward to extend our discussion to higher rank masses. We call $\tilde{T}$ the resulting theory, which still has the same quiver diagram of $T$, but with reduced ranks for some of the global nodes, as we now explain. The mass term clearly breaks the global symmetry down to $(\tilde{N}_1,\tilde{N}_2,\tilde{N}_3,\tilde{N}_4)=(N_1-1,N_2-1,N_3,N_4)$.\footnote{In fact there is an additional global $U(1)$. We can capture it by thinking that the massive flavors are connected to a new, rank 1, global node.} The gauge group is unaffected so its rank is $\tilde{N}_0=N_0$.

We now take $\tilde{T}$ as the new starting point. Acting with quadrality, we obtain a {\it deformed sequence} of theories $\tilde{T}$, $\tilde{T}'$, $\tilde{T}''$ and $\tilde{T}'''$. Once again, the quivers for the deformed sequence are those in \fref{quadrality_periodicity}. The differences with respect to the original sequence are the following. The ranks of the flavor nodes are $(N_1-1,N_2-1,N_3,N_4)$ for all the theories. In addition, using the initial $\tilde{N}_i$'s in \eref{gauge_group_rank_quadrality} we determine the evolution of the rank of the gauge group is
\beq
\begin{array}{c|c|c|c}
\ \ \ \ \ \ \tilde{T} \ \ \ \ \ \ & \ \ \ \ \ \  \tilde{T}' \ \ \ \ \ \ & \ \ \ \ \ \ \tilde{T}'' \ \ \ \ \ \ & \ \ \ \ \ \ \tilde{T}''' \ \ \ \ \ \ \\ \hline
N_0 & N_0'-1 & N_0'' & N_0 '''
\end{array}
\label{N_0_deformation_a}
\eeq

\smallskip

%=================================================================
\noindent{\bf b) \underline{$\Lambda_{30}X_{04}$ mass}} 
%=================================================================

\smallskip

This deformation is very similar to the one we have just considered. It breaks the global symmetry down to $(N_1,N_2,N_3-1,N_4-1)$. The rank of the gauge group is unaffected in the initial theory and evolves as follows
\beq
\begin{array}{c|c|c|c}
\ \ \ \ \ \ \tilde{T} \ \ \ \ \ \ & \ \ \ \ \ \  \tilde{T}' \ \ \ \ \ \ & \ \ \ \ \ \ \tilde{T}'' \ \ \ \ \ \ & \ \ \ \ \ \ \tilde{T}''' \ \ \ \ \ \ \\ \hline
N_0 & N_0' & N_0'' & N_0 '''-1
\end{array}
\eeq

\smallskip

%=================================================================
\noindent{\bf c) \underline{$\Lambda_{30}\Lambda_{02}$ mass}} 
%=================================================================

\smallskip

Finally, we can deform $T$ by introducing a $\Lambda_{30}\Lambda_{02}$ mass, namely $H_{\Lambda_{30},\Lambda_{02}}=m$. The global symmetry is reduced to $(N_1,N_2-1,N_3-1,N_4)$. The gauge group remains the same in the starting theory and then follows the sequence
\beq
\begin{array}{c|c|c|c}
\ \ \ \ \ \ \tilde{T} \ \ \ \ \ \ & \ \ \ \ \ \  \tilde{T}' \ \ \ \ \ \ & \ \ \ \ \ \ \tilde{T}'' \ \ \ \ \ \ & \ \ \ \ \ \ \tilde{T}''' \ \ \ \ \ \ \\ \hline
N_0 & N_0' & N_0''-1 & N_0 '''
\end{array}
\eeq

\medskip

It is interesting to notice that a common feature of the three deformations considered above is that, for each of them, only one theory in the sequence gets higgsed.\footnote{In an abuse of language, what we mean by higgsing is that some of the gauginos become massive, which reduces the gauge symmetry.}

%=================================================================
\paragraph{Connecting the Original and the Deformed Sequences.}
%=================================================================

Alternatively, if the quadrality proposal is correct, we should be able to obtain the deformed sequences by mapping the original deformations of $T$ to deformations of $T'$, $T''$ and $T'''$. As we will now explain, it is possible to verify that this is indeed the case.

Before doing so, it is convenient to discuss how general deformations are mapped under quadrality. In the context of $\mathcal{N}=1$ matrix models, we refer to any modification of $J$- or $H$-terms as a deformation. Deformations modify the global symmetry of a theory, while preserving its gauge symmetry and matter content.  When moving to a dual theory we must now include all interactions that are consistent with the new global symmetry.  The modification in the interactions of the dual theory is identified with the map of the original deformation. Typically, we are interested in introducing new interaction terms to the original theory, which reduces the global symmetry. As a result, the dual theory admits new interaction terms that are the translation of the deformation. This prescription is very general and is the one used when mapping deformations under dualities in other dimensions. For example, it is precisely the approach one uses in the well-known case of mapping superpotential deformations of 4d $\mathcal{N}=1$ theories under Seiberg duality.\footnote{It is important to note that while it is relatively straightforward to map deformations under duality with our prescription, this cannot be done by simply rewriting the component expansions of some terms in the Lagrangian. Here, again, $4d$ Seiberg duality provides a familiar example. Deformations can be easily implemented at the level of the superpotential, just replacing combinations of chiral fields that are charged under the dualized gauge group by mesons. This process, however, cannot be implemented as a reorganization of the component expansion of the Lagrangian.}

For concreteness, let us focus on case $(a)$, namely on a rank 1 $X_{10}\Lambda_{02}$ mass term. The other deformations can be understood using similar ideas. The first dual theory, $T'$, contains a Fermi meson $\mu_{12}=X_{10}\Lambda_{02}$. The deformation of $T$ given by \eref{deformation_a} maps to an additional constant contribution to $J_{\mu_{12}}$ as follows
\beq
T: J_{\Lambda_{02}}= m \, \phi_{10} \ \ \ \ \to \ \ \ \  T': J_{\mu_{12}}=X_{20}X_{01}+m \, ,
\eeq
where the first term in $J_{\mu_{12}}$ is the usual coupling between the Fermi meson $\mu_{12}$ and the chiral flavors in $T'$. The constant term in $J_{\mu_{12}}$, where $m$ is the original rank 1 mass matrix, is precisely the new interaction that is allowed when the global symmetry is reduced to $(N_1-1,N_2-1,N_3,N_4)$.\footnote{This is the $0d$ analogue of the map, at the level of the superpotential, between a mass term for quarks in $4d$ SQCD and a linear term for mesons in its Seiberg dual.} 

The action now contains the term
\beq
S_J =\bar{J}_{\mu_{12}}J_{\mu_{12}}+\ldots= |\phi_{20}\phi_{01}+m|^2 + \ldots \, ,
\eeq
which fixes at a non-zero value an entry in each of the diagonalized $\phi_{20}$ and $\phi_{01}$. This, in turn, reduces the rank of the gauge group to $N_0-1$, in perfect agreement with \eref{N_0_deformation_a}.

\begin{figure}[htbp]
	\centering
	\includegraphics[width=9.5cm]{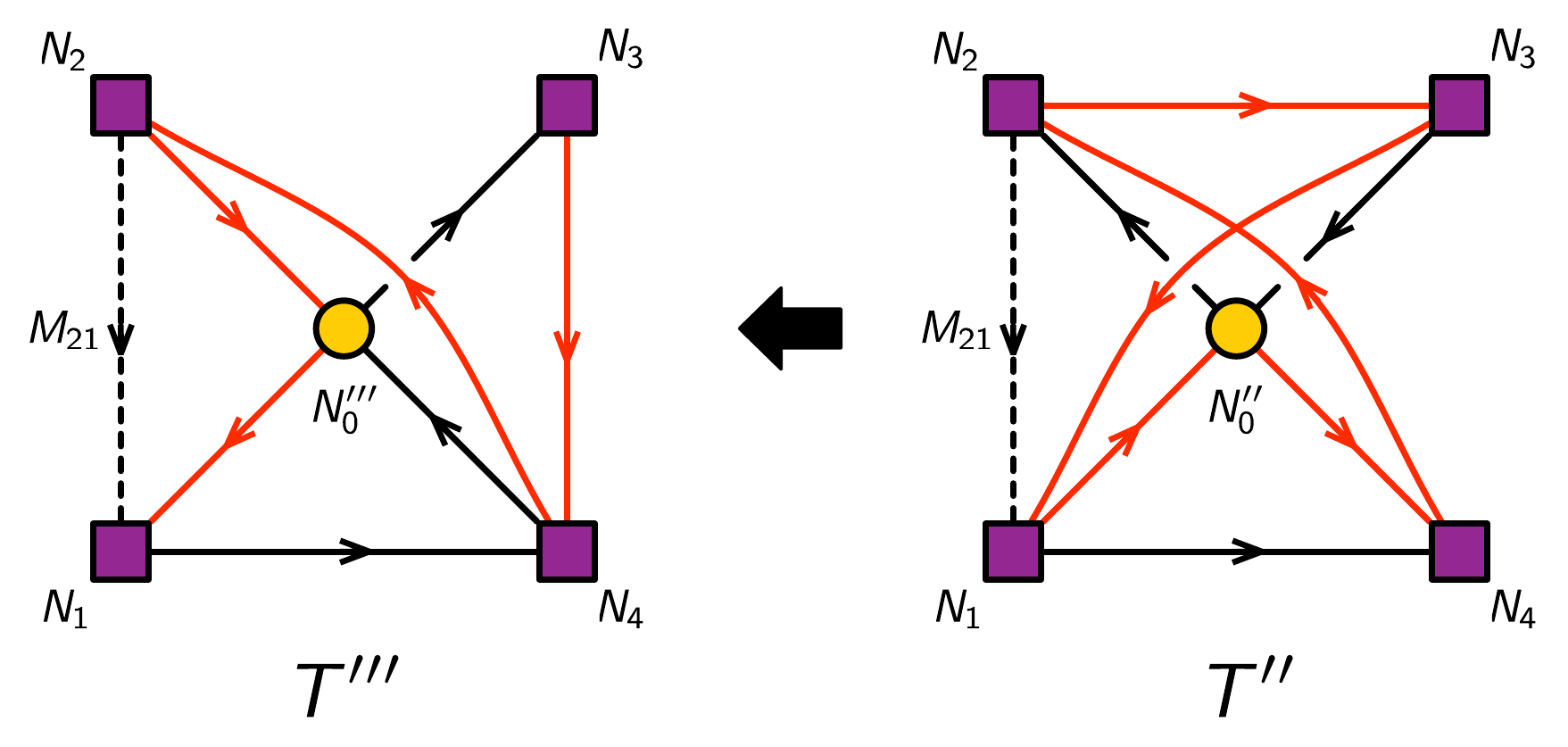}
\caption{The two last steps in the quadrality sequence, $T''$ and $T'''$. We show the position of the meson $M_{21}$, which becomes massive by coupling to other fields and is integrated out.}
	\label{M21_T''_T'''}
\end{figure}

Having correctly reproduced $\tilde{T}'$ by mapping the deformation, let us now consider the two remaining theories in the sequence. The fixed non-zero values for $\phi_{20}$ and $\phi_{01}$ in $T'$ map to mass terms in both $T''$ and $T'''$ that break the global symmetry down to $(N_1-1,N_2-1,N_3,N_4)$, as expected. The two mass terms are:
\beq
\begin{array}{cccrcl}
T'' \to \tilde{T}'' & \ \ \ & J\mbox{-term:} & J_{\Lambda_{10}}& =& m \, \phi_{02} \\[.15cm]
T''' \to \tilde{T}''' & & H\mbox{-term:} & \ H_{\Lambda_{20}, \Lambda_{01}}& =& m 
\end{array}
\label{mass_deformations_T''_T'''}
\eeq
We can understand these deformations as follows. When $T'$ is dualized, a chiral meson $M_{21}=X_{20} X_{01}$ is generated. This meson is not shown in phases $T''$ and $T'''$ of \fref{quadrality_periodicity} because it becomes massive and can be integrated out. \fref{M21_T''_T'''} shows where $M_{21}$ would be in $T''$ and $T'''$. This picture is just intended as a visual reference; to reintroduce $M_{21}$ in these theories we should also integrate in the fields they paired with.

%======================================================================
%======================================================================

These theories contain the following interactions:
\beq
\begin{array}{cccrcl}
T'' & \ \ \ & J\mbox{-term:} & J_{\Lambda_{10}}& =& \phi_{02} \,\phi_{M_{21}} \\[.15cm]
T'''& & H\mbox{-term:} & \ H_{\Lambda_{20}, \Lambda_{01}}& =& \phi_{M_{21}}
\end{array}
\label{M21_couplings_T''_T'''}
\eeq
The non-zero values for $\phi_{20}$ and $\phi_{01}$ in $T'$ that we discussed above, translate into a non-zero value for $\phi_{M_{21}}$. When plugged into \eref{M21_couplings_T''_T'''}, it produces the mass deformations in \eref{mass_deformations_T''_T'''}. In addition, no flavor gets a non-zero value in either $T''$ or $T'''$ so the gauge group is not higgsed. This is in agreement with the fact that the ranks of the gauge group in $\tilde{T}''$ and $\tilde{T}'''$ are equal to those in the undeformed sequence. 

In summary, we have correctly reproduced the entire deformed sequence. This matching provides a rather non-trivial check of the quadrality proposal, which takes into account all types of mesons and interaction terms.

%=================================================================
\subsection{Chiral Ring \label{sec:chiral-ring}}
%=================================================================

As another check of quadrality, we compare the chiral ring of the four theories. For simplicity, we identify a few elements of the chiral ring, focusing on operators that can be expressed as superfields.
We restrict to operators built out of a small number of matter fields, leaving a complete enumeration for future work. 

The chiral ring is given by cohomology classes of $Q$, i.e. it consists of gauge invariant operators that are $Q$-closed but not $Q$-exact. The top components of superfields are, by definition, $Q$-exact. The only exception is $\overline{X}$, which has a single component.\footnote{As mentioned earlier, even though we sometimes consider $X$ and $\overline{X}$ separately, we should keep in mind that they combine into a single superfield.} We are thus interested in the case in which only the lowest component of a gauge invariant superfield, either elementary or a product, survives. Such a component then becomes an element of the chiral ring. The chiral ring is determined on-shell which, among other things, requires that $G_a=\bar{J}_a$ for all elementary Fermi fields. 

Let us first consider gauge invariant elementary fields. For quick reference, below we summarize some of the superfield discussion of section \sref{sec:super-matrix} and indicate whether the lowest components of the operators are in the chiral ring.
 \beq
\begin{array}{|c|c|c|c|}
\hline
\ \ \ \mbox{Superfield} \ \ \ & \ \ \ \mbox{lowest} \ \ \ & \ \ \ Q(\mbox{lowest}) \ \ \ & \ \ \ \mbox{Chiral ring?} \ \ \ \\ \hline
X & \phi & \psi & \times \\ \hline 
\overline{X} & \bar{\phi} & 0 & \checkmark \\ \hline 
\Lambda & \lambda & G & \circ \\ \hline 
\end{array}
\label{chiral_ring_1_field}
\eeq
where we have separated the $X$ and $\overline{X}$ contributions.

We can repeat the exercise for products of two matter fields. Following \eref{product_1}-\eref{product_3}, we obtain
\beq
\begin{array}{|c|c|c|c|}
\hline
\ \ \ \mbox{Superfield} \ \ \ & \ \ \ \mbox{lowest} \ \ \ & \ \ \ Q(\mbox{lowest}) \ \ \ & \ \ \ \mbox{Chiral ring?} \ \ \ \\ \hline
X_1 X_2 & \phi_1 \phi_2 & \phi_1 \psi_2+\psi_1 \phi_2 & \times \\ \hline 
\overline{X}_1 \overline{X}_2 & \bar{\phi}_1 \bar{\phi}_2 & 0 & \checkmark \\ \hline 
X_1 \Lambda_2 & \phi_1 \lambda_2 & \phi_1 G_2 + \psi_1 \lambda_2 & \times \\ \hline
\overline{X}_1  \Lambda_2 & \bar{\phi}_1 \lambda_2 & \bar{\phi}_1 G_2 & \circ \\ \hline
\end{array}
\label{chiral_ring_2_fields}
\eeq
The ``$\circ$" category in \eref{chiral_ring_1_field} and \eref{chiral_ring_2_fields} indicates operators that become elements of the chiral ring if the $G$-components of the corresponding Fermi fields vanish. This is automatically the case if such Fermis do not participate in any $J$-term oriented loop in the quiver.

Below we list operators in the chiral ring for the quadrality sequence shown in \fref{quadrality_periodicity}. For convenience, we refer to the operators in terms of superfields, with the understanding that their lowest components are the objects of interest.

%=================================================================
\beq
\begin{array}{|c|c|c|c|c|}
\hline
 & T & T' & T'' & T''' \\ \hline
\ \mathcal{O}_{41} \ \ &  \overline{X}_{04}  \overline{X}_{10}                                     & \overline{M}_{14} & \overline{M}_{14}  & \overline{M}_{14} \\ \hline
\ \mathcal{O}_{31} \ \ & \Lambda_{30} \overline{X}_{10}         & \mu_{31} & \mu_{31} & \ \ \overline{X}_{03} \Lambda_{01}-\mu_{34} \overline{M}_{14} \ \ \\ \hline
\ \mathcal{O}_{42} \ \ & \overline{X}_{04} \Lambda_{02}         & \ \ \Lambda_{40} \overline{X}_{20} - \overline{M}_{14} \mu_{12} \ \ &  \mu_{42} & \mu_{42} \\ \hline
\end{array}
\label{chiral_ring_quadrality_sequence}
\eeq
%=================================================================

\smallskip

%=================================================================
\paragraph{Matching $T$ and $T'$.}
%=================================================================

In order to illustrate the main ideas that go into the matching, it is instructive to discuss the correspondence between $T$ and $T'$ in detail. The rest of the theories follow a similar logic. In particular the analysis of $T'''$ is identical to the one of $T'$ up to a reflection with respect to a vertical axis. Let us discuss the rows in \eref{chiral_ring_quadrality_sequence} that deserve special comments.
\begin{itemize}

\item The operator $\mathcal{O}_{31}$ is given by $\lambda_{30}\bar{\phi}_{10}$ in $T$ and by  $\lambda_{\mu_{31}}$ in $T'$. It may appear that $T'$ also contains the operator $\bar{\phi}_{01} \lambda_{03}$ between this pair of nodes. However, the theory has $H_{\Lambda_{03},\mu_{31}}=\bar{\phi}_{01}$, which gives rise to the coupling $\lambda_{03} \lambda_{\mu_{31}} \bar{\phi}_{01}$. Since $\mu_{31}$ does not participate in any other loop in the quiver, the equation of motion for $\lambda_{\mu_{31}}$ forces $\bar{\phi}_{01} \lambda_{03}$ to vanish on-shell.

\item $\mathcal{O}_{42}$ corresponds to $\bar{\phi}_{04} \lambda_{02}$ in $T$. Interestingly, there are two operators in $T'$ with the right properties: $\lambda_{40} \bar{\phi}_{02}$ and $\bar{\phi}_{{M}_{14}} \lambda_{\mu_{12}}$. Which one should we use? First of all, the two operators are not related by an equation of motion, so we cannot restrict to just one of them. Remarkably, for $Q$-closedness it is necessary to consider a linear combination of them: $\lambda_{40} \bar{\phi}_{02}-\bar{\phi}_{{M}_{14}} \lambda_{\mu_{12}}$. We have
\beq
\begin{array}{rcl}
\{Q, \lambda_{40} \bar{\phi}_{02}-\bar{\phi}_{{M}_{14}} \lambda_{\mu_{12}} \} & = & G_{40} \bar{\phi}_{20} - \bar{\phi}_{{M}_{14}} G_{\mu_{12}} \\[.1cm]
& \simeq  & \bar{\phi}_{14} \bar{\phi}_{10} \bar{\phi}_{20} - \bar{\phi}_{14} \bar{\phi}_{10} \bar{\phi}_{20} = 0 \, .
\end{array}
\eeq
Both $\Lambda_{40}$ and $\mu_{12}$ participate in $J$-term loops, so their respective $G$-components do not automatically vanish, but they compensate when combined. A similar explanation applies to $\mathcal{O}_{31}$ in $T'''$.

\end{itemize}

\medskip

%=================================================================
\paragraph{Additional Comments.}
%=================================================================

Further things that work nicely for all the theories are:

\begin{itemize}

%=================================================================
\item{\bf Elementary singlet Fermis.}
%=================================================================
Notice that the three theories other than $T$ have singlet Fermi fields: $\mu_{12}$ in $T'$, $\mu_{23}$ in $T''$ and $\mu_{34}$ in $T'''$. They are not in the chiral ring, because they all participate in $J$-term loops and hence their $G$-components do not vanish. This is good, since there are no matching operators in the other theories, most notably in $T$.

%=================================================================
\item{\bf $X\Lambda$ Fermis.}
%=================================================================
There are two types of composite Fermi fields, which are of the general forms $X\Lambda$ and $\overline{X}\Lambda$. In \eref{chiral_ring_2_fields} we argued that the $X\Lambda$ operators are not in the chiral ring. Indeed, such operators do not match between different theories. A simple example is $X_{20}\Lambda_{03}$ in $T'$, which does not have a counterpart in $T$.

\end{itemize}

Chiral rings can be fairly non-trivial in theories with multiple matter fields and gauge groups, even if restricted to special subsectors. This point is illustrated in \sref{scoordring}, where we compute all the chiral ring operators that only consist of chiral fields for two quadrality dual theories on D(-1)-branes probing local $\mathbb{CP}^4$.

%=================================================================
\section{Quadrality Networks}
%=================================================================

\label{section_quadrality_networks}

Theories connected by sequences of quadrality transformations can be nicely organized in terms of {\it quadrality networks}. Analogous constructions exist for Seiberg duality \cite{Cachazo:2001sg,Franco:2003ja,Franco:2004jz} and triality \cite{Gadde:2013lxa,Franco:2016nwv}. These networks become particularly interesting for theories with multiple gauge groups. 

\begin{figure}[H]
	\centering
	\includegraphics[height=7.5cm]{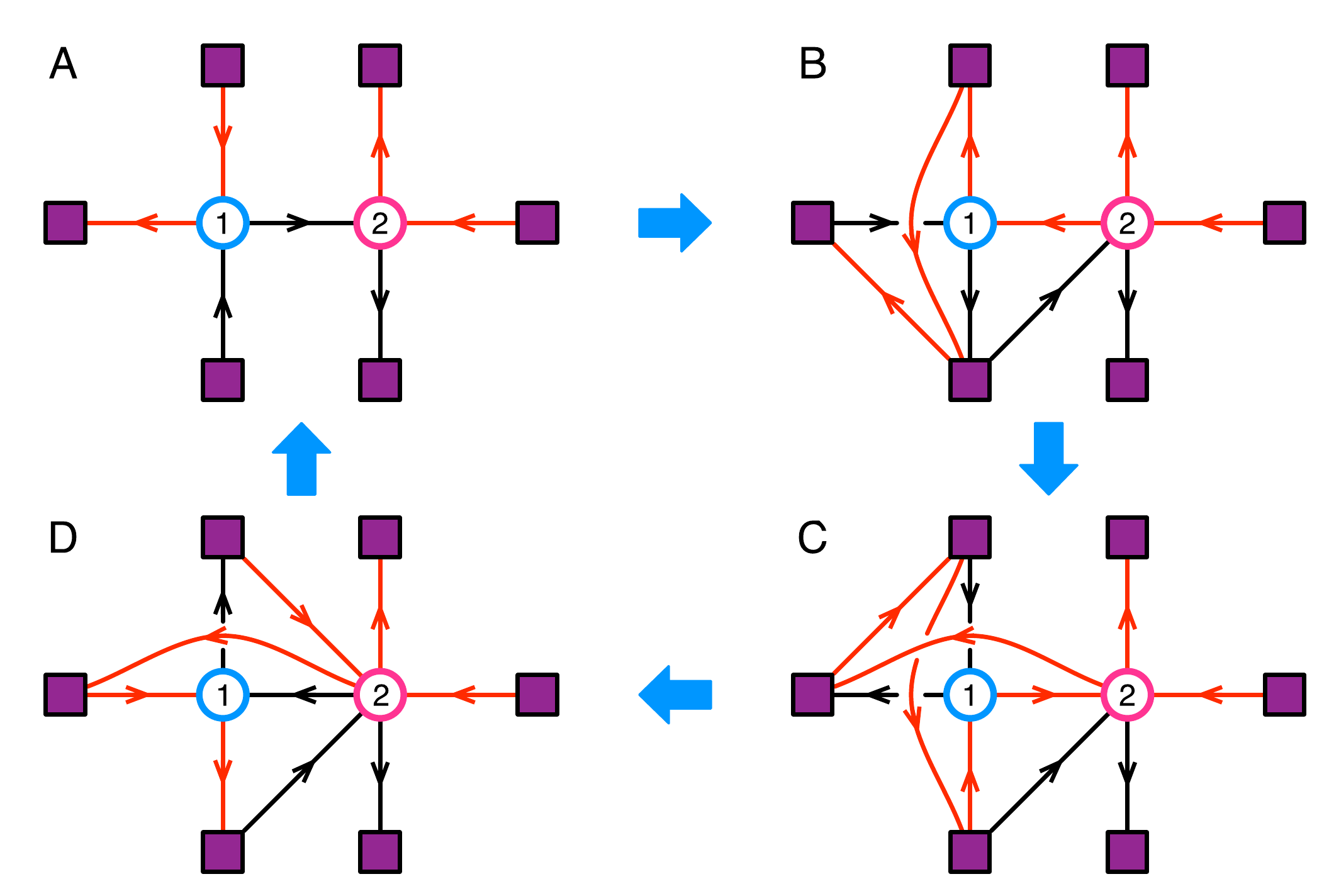}
\caption{Quadrality loop obtained by acting repeatedly with quadrality on the same gauge node, node 1, of a 2-gauge/6-flavor quiver.}
	\label{quad-2-ex}
\end{figure}

One simple generalization of the simple SQCD-like theory considered in previous sections consists of merging several copies of the basic 1-gauge/4-flavor quiver to make up $n$-gauge/$(2n+2)$-flavor quivers. An $n=2$ example with the quadrality action on gauge node 1 is shown in \fref{quad-2-ex}. 

%=================================================================
%=================================================================

\begin{figure}[htbp]
	\centering
	\includegraphics[height=14cm]{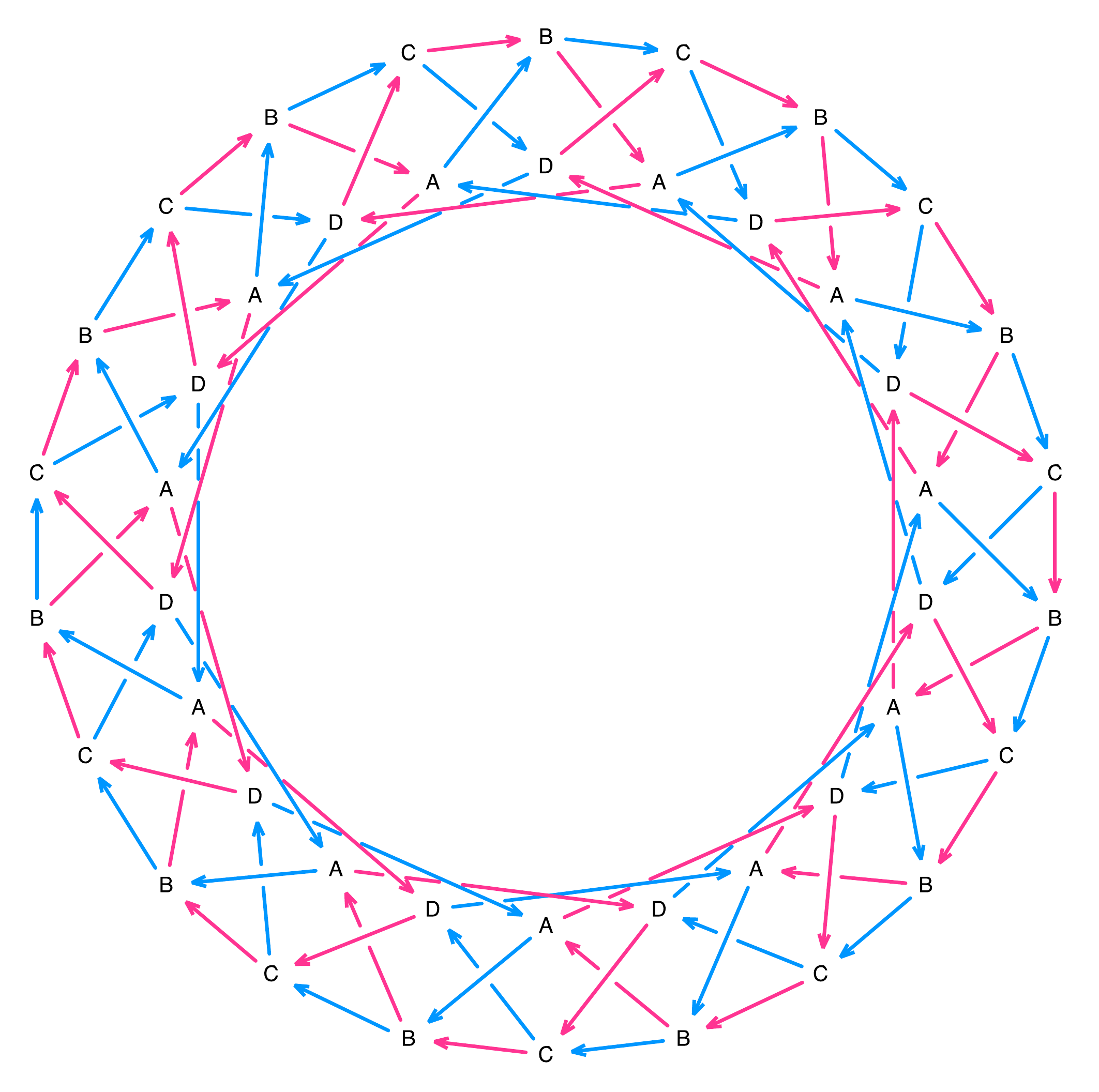}
\caption{The quadrality network for 2-gauge/6-flavor quivers}
	\label{quad-2-gauge}
\end{figure}

The quadrality actions on the two gauge nodes do not commute, so all possible combinations of quadrality lead to a network of quivers. It turns out that there are 44 quivers at $n=2$. The quivers can be divided into four types according to the type of field connecting the two gauge nodes. Quivers of the same type may have different meson contents. \fref{quad-2-gauge} shows the complete quadrality network of $n=2$ quivers. In the figure, the labels A, B, C, D denote the four types of quivers as in \fref{quad-2-ex}. The blue and pink arrows indicate quadrality transformations on node 1 and 2, respectively. The length-4 closed oriented loops consisting of arrows of a given color correspond to four consecutive quadrality transformations on the same node. As it occurs for Seiberg duality and triality, more general quivers can lead to infinite quadrality networks.

%=================================================================
%=================================================================

\newpage
%=================================================================
\section{D-Brane Theories}
%=================================================================

\label{section_D-brane_examples}

In this section we study $\mathcal{N}=1$ matrix models arising on the worldvolume of D(-1)-branes probing toric CY 5-folds.

%=================================================================
\subsection{$\mathbb{C}^5$}
%=================================================================

Let us first consider D(-1)-branes in flat $10d$ spacetime, i.e. $\mathbb{C}^5$. This theory is often called the {\it Type IIB matrix model} and has been proposed as a nonperturbative formulation of type IIB string theory \cite{Ishibashi:1996xs,Fukuma:1997en,Aoki:1998vn}. The models presented in this and the coming section illustrate how the general structures discussed in section \sref{sec:super-matrix} arise in D-brane constructions.

Bosons and fermions are decomposed in terms of the $SU(5) \times U(1) \subset SO(10)$ global symmetry as follows
\begin{align}
\begin{split}
\Phi &: {\bf 10_v} \rightarrow \phi^m({\bf 5}_1) + \bar{\phi}_m({\bf \bar{5}}_{-1}) 
\\
\Psi &: {\bf 16_s} \rightarrow \chi(1_{5/2}) + \lambda_{mn} ({\bf \overline{10}}_{1/2}) + \psi^m({\bf 5}_{-3/2})
\end{split}
\end{align}
Here $m$ and $n$ are indices in the fundamental representation of $SU(5)$. We choose the convention where complex conjugation exchanges superscripts and subscripts. Contraction of conjugate indices ($A^m B_m$) implies summation.

The 16 supercharges decompose as
\begin{align}
Q &: {\bf 16_c} \rightarrow {\bf 1}_{-5/2} + {\bf 10}_{-1/2} + {\bf \bar{5}}_{3/2} \, .
\label{Q-decomp}
\end{align}
We will only use the singlet supercharge. 

The on-shell form of the action of the matrix model can be split into three parts:
\begin{align}
S = S_D + S_J + S_H \,.
\end{align}
The $D$-term is
\begin{align}
S_D = \Tr \left( \frac{1}{2}[ \phi^m , \bar{\phi}_m]^2 + \chi [ \bar{\phi}_m , \psi^m ]  \right) \,.
\label{C5-SD}
\end{align}
The $J$-term is 
\begin{align}
S_J = \Tr \left(\frac{1}{2} [ \phi^m , \phi^n][\bar{\phi}_n , \bar{\phi}_m ]+ \lambda_{mn} [ \phi^m , \psi^n ] \right) \,.
\label{C5-SF}
\end{align}
The $H$-term is 
\begin{align}
S_H = \frac{1}{8}\varepsilon^{mnpqr} \Tr \left( \lambda_{mn} \lambda_{pq} \bar{\phi}_r \right) \,, 
\label{C5-SH}
\end{align}
where $\varepsilon^{mnpqr}$ is the $SU(5)$ invariant tensor. 

The supersymmetry variation is 
\begin{align}
\begin{split}
\delta \phi^m &= \epsilon \psi^m \,, 
\quad \delta \psi^m = 0 \,,
\quad 
\delta \bar{\phi}_m = 0 \,, 
\\
\delta \lambda_{mn} &= [\bar{\phi}_m, \bar{\phi}_n] \epsilon \,,
\quad 
\delta \chi = [\phi^m , \bar{\phi}_m] \epsilon \,.
\\
\end{split}
\label{C5-susy}
\end{align}
It is easy to show that $S_D$, $S_J$ and $S_H$ are separately invariant under a supersymmetry transformation. We can introduce independent coupling constants for the three terms. Ratios among the three couplings are fixed only if we turn on non-singlet supercharges from \eqref{Q-decomp}. 

Unlike $S_D$ or $S_J$, the supersymmetry of $S_H$ does not rely on a cancellation between a purely bosonic term and a fermion bilinear term. 
Explicitly, 
\begin{align}
\begin{split}
8 (\delta S_H) &= \epsilon^{mnpqr} \epsilon \,\Tr\left([\bar{\phi}_m, \bar{\phi}_n] \lambda_{pq} \bar{\phi}_r - \lambda_{mn} [\bar{\phi}_p, \bar{\phi}_q] \bar{\phi}_r \right) 
\\
&= - \epsilon^{mnpqr} \epsilon \,\Tr\left(\lambda_{mn} [[\bar{\phi}_p, \bar{\phi}_q], \bar{\phi}_r ]\right) = 0 \,.
\end{split}
\end{align}
In the last step, we used the Jacobi identity. 

It is straightforward to rephrase this theory in the (mostly) off-shell formalism of section \sref{sec:super-matrix}. 
We simply note that
\begin{align}
J^{mn}(\phi) = - [\phi^m, \phi^n] \,,
\quad 
\bar{J}_{mn}(\bar{\phi}) = [\bar{\phi}_m,\bar{\phi}_n] \,,
\quad 
\overline{H}^{mn,pq}(\bar{\phi}) = \varepsilon^{mnpqr} \bar{\phi}_r
\,.
\end{align}

%=================================================================
\subsection{Local $\mathbb{CP}^4$}
%=================================================================

\label{section_C5/Z5_(1,1,1,1,1)}

Let us now consider D(-1)-branes probing local $\mathbb{CP}^4$, namely the $\mathbb{C}^5/\mathbb{Z}_5$ orbifold with action $(1,1,1,1,1)$. This geometry is the simplest toric CY$_5$ that can be studied using mirror symmetry and, as discussed in section \sref{section_local_geometry_order_n_dualities}, it is the minimal local geometry realizing quadrality. 

\begin{figure}[htbp]
	\centering	
	\resizebox{0.55\hsize}{!}{
	\includegraphics[height=5cm]{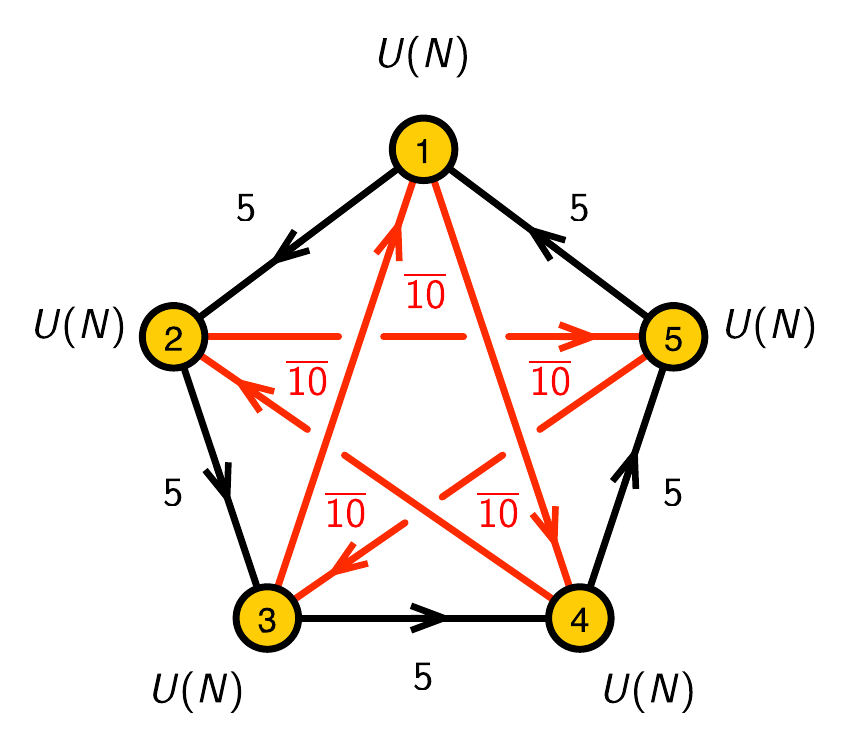}
	}
\caption{Quiver diagram for local $\mathbb{CP}^4$.}
	\label{fig:C5Z5-quiver}
\end{figure}

The corresponding gauge theory is obtained from the one for $\mathbb{C}^5$ presented in the previous section by standard orbifold techniques \cite{Douglas:1996sw,Douglas:1997de}. Its quiver diagram is shown in \fref{fig:C5Z5-quiver}. We have chiral multiplets $X^m_{i,i+1}$, Fermi multiplets $\Lambda_{mn}^{i+2,i}$ and gaugino multiplets $\chi_{i,i}$.

%=================================================================
%=================================================================

The on-shell supersymmetry variation is a natural generalization of \eqref{C5-susy}:
\begin{align}
\begin{split}
\delta \phi^m_{i,i+1} &= \epsilon \psi^m_{i,i+1} \,, 
\quad 
\quad \delta \psi^m_{i,i+1} = 0 \,,
\quad 
\delta \bar{\phi}_m^{i+1,i} = 0 \,, 
\\
\delta \lambda_{mn}^{i+2,i} &= (\bar{\phi}_m^{i+2,i+1} \bar{\phi}_n^{i+1,i} - \bar{\phi}_n^{i+2,i+1} \bar{\phi}_m^{i+1,i} )  \epsilon \,,
\\
\delta \chi_{i,i} &= (\phi^m_{i,i+1} \bar{\phi}_m^{i+1,i} - \bar{\phi}_m^{i,i-1}\phi^m_{i-1,i}) \epsilon \,.
\end{split}
\label{C5Z5-susy}
\end{align}
The terms in the action are 
\begin{align}
S_D &= \sum_{i=1}^5 \Tr \left[ \frac{1}{2}|X_{i,i+1}^m \bar{\phi}_m^{i+1,i} - \bar{\phi}_m^{i,i-1} X_{i-1,i}^m |^2 
+ \chi_{i,i} (\bar{\phi}_m^{i,i-1}\psi^m_{i-1,i} - \psi^m_{i,i+1}\bar{\phi}_m^{i+1,i}) 
\right]\,,
\nn \\
S_J &=  \sum_{i=1}^5 \Tr \left[\frac{1}{2} |\phi^m_{i-1,i} \phi^n_{i,i+1} - \phi^n_{i-1,i} \phi^m_{i,i+1}|^2 
+ \lambda_{mn}^{i+1,i-1} (\phi^m_{i-1,i} \psi^n_{i,i+1} - \psi^n_{i-1,i} \phi^m_{i,i+1}) 
\right] \,,
\nn \\
S_H &= \frac{1}{8} \varepsilon^{mnpqr} \sum_{i=1}^5 \Tr\left[ \lambda_{mn}^{i+2,i} \lambda_{pq}^{i,i-2} \bar{\phi}_r^{i-2,i+2}
\right] \, ,
\label{action_C5/Z5}
\end{align}
where the quiver node indices are defined (mod 5). The absolute-value-square of a complex matrix is defined as $|A|^2 = A A^\dagger$. Once again, $S_D$, $S_J$ and $S_H$ are separately invariant under a supersymmetry transformation.

%=================================================================
\subsection{The Chiral Ring and the Probed CY$_5$ \label{scoordring}} 
%=================================================================

For theories on D(-1)-branes, the chiral ring operators consisting exclusively of chiral fields should reproduce the coordinate ring of the probed CY$_5$. More precisely, following the discussion in section \sref{sec:chiral-ring}, the operators we are interested in are products of $\bar{\phi}$ components of chiral fields. For simplicity, since we are focusing only on chiral fields, we can drop the conjugation in our discussion. Using local $\mathbb{CP}^4$ as an example, in this section we show that the CY$_5$ can be recovered from two different quadrality phases. This fact simultaneously demonstrates the connection of the chiral ring to the probed CY$_5$ and the invariance of this particular sector under quadrality.

%=================================================================
\subsubsection*{The Original Theory}
%=================================================================

Let us first consider the theory presented in section \sref{section_C5/Z5_(1,1,1,1,1)}. For simplicity, let us focus on a single D(-1)-brane, namely we set $N=1$ in the quiver of \fref{fig:C5Z5-quiver}. For generic $N$, we can diagonalize all fields and the full answer is the $N^{th}$ symmetric product of the $N=1$ result. We can construct the following $5^5$ gauge invariant operators
\beal{es100}
M^{m_1 m_2 m_3 m_4 m_5} = \phi^{m_1}_{12} \phi^{m_2}_{23} \phi^{m_3}_{34} \phi^{m_4}_{45} \phi^{m_5}_{51}\, , \qquad m_i=1,\ldots,5 \,.
\eea
In terms of the $SU(5)$ global symmetry, they decompose as follows
\beq
\begin{array}{rcl}
{\bf 5} \times {\bf 5} \times {\bf 5} \times {\bf 5} \times {\bf 5} & = & {\bf 1}+4({\bf 24})+5({\bf 75})+6({\bf 126})+{\bf \underline{126}'}+5({\bf 175}')+4({\bf 224}) \\[.1cm]
& = & [0,0,0,0]+4[1,0,0,1] +5[0,1,1,0]+6[2,0,1,0]+\underline{[5,0,0,0]}\\[.1cm]
& & +5[1,2,0,0]+4[3,1,0,0] \, ,
\end{array}
\label{SU(5)_reps_GIOs}
\eeq
where, in order to distinguish representations with equal dimension we also provided the corresponding Dynkin labels.

Vanishing of the bosonic potential in \eref{action_C5/Z5} gives rise to the following relations
\beal{es101}
\phi^m_{i-1,i} \phi^n_{i,i+1} - \phi^n_{i-1,i} \phi^m_{i,i+1} = 0 \, ,
\eea
which fully symmetrize the indices in $M^{m_1 m_2 m_3 m_4 m_5}$. Accordingly, only the ${\bf 126}'=[5,0,0,0]$, i.e. the totally symmetric 5-index representation of $SU(5)$, survives from the gauge invariant operators in \eref{SU(5)_reps_GIOs}. We call them the {\it generators} of the chiral ring and label them $M_s^{m_1 m_2 m_3 m_4 m_5}$, with $1\leq m_1 \leq m_2 \leq m_3 \leq m_4 \leq m_5 \leq 5$.

The generators satisfy first order quadratic relations. First, note that
\beq
\begin{array}{rcl}
{\bf 126}' \times {\bf 126}' & = & {\bf 1004}+{\bf 2574}+{\bf 3850}'+{\bf 4125}+{\bf 3150}+{\bf 1176}'  \\[.1cm]
& = &[10,0,0,0] + [8,1,0,0] + [6,2,0,0] + [4,3,0,0] \\[.1cm]
& & + [2,4,0,0]+ [0,5,0,0] \, .
\end{array}
\label{es102}
\eeq
The $J$-term relations \eref{es101} imply that in a product $M_s^{m_1 m_2 m_3 m_4 m_5} M_s^{n_1 n_2 n_3 n_4 n_5}$ any pair of indices $(m_i,n_i)$ can be swapped leaving the product invariant. Using this, it is possible to show that the generators obey $7000$ quadratic relations, which transform in the ${\bf 3850}'+{\bf 3150}$ representation of $SU(5)$. They can be written explicitly as follows
\beq
\begin{array}{cccl}
{\bf 3850}' & = & [6,2,0,0]: & R^{ijklmn}_{pqrstu} = M_s^{ijk v_1 v_2} M_s^{lmn w_1 w_2} \epsilon_{v_1 w_1 pqr} \epsilon_{v_2 w_2 stu} = 0 \, ,~ \\[.2 cm]
{\bf 3150}  & = & [2,4,0,0]: & T^{ij}_{p_1 p_2 p_3 q_1 q_2 q_3 r_1 r_2 r_3 s_1 s_2 s_3} =  M_s^{i m_1 m_2 m_3 m_4} M_s^{j n_1 n_2 n_3 n_4} \\
& &  & \hspace{1cm}
\epsilon_{m_1 n_1 p_1 p_2 p_3}
\epsilon_{m_2 n_2 q_1 q_2 q_3}
\epsilon_{m_3 n_3 r_1 r_2 r_3}
\epsilon_{m_4 n_4 s_1 s_2 s_3} = 0
\, .~
\end{array}
\label{es103}
\eeq
The variety formed by the generators subject to their first order relations is not a complete intersection. The plethystic logarithm of the Hilbert series $g(t,x_i; \mathbb{C}^5/\mathbb{Z}_5)$ of the variety is thus not finite and takes the form
\beal{es105}
\text{PL}[g(t,x_i; \mathbb{C}^5/\mathbb{Z}_5)] = [5,0,0,0] t - ([6,2,0,0]+[2,4,0,0]) t^2 + \dots \, ,~
\eea
where $t$ is a fugacity counting the degree in terms of the generators $M_s^{m_1 m_2 m_3 m_4 m_5}$. The previous analysis is in precise agreement with the computation of the Hilbert series directly from the toric geometry of $\mathbb{C}^5/\mathbb{Z}_5$, as explained in appendix \sref{sapphs}. We hence conclude the chiral ring reproduces the coordinate ring of the probed CY$_5$. The $SU(5)$ charges of the generators transforming in the ${\bf 126}'$ representation form a convex polytope in $\mathbb{Z}^4$. This polytope -- the lattice of generators -- is the dual of the toric diagram of $\mathbb{C}^5/\mathbb{Z}_5$. Such relation between the generators and the polytope dual to the toric diagram is described in detail for the simpler example of $\mathbb{C}^3/\mathbb{Z}_3$ in appendix \sref{sappc3z3}.

%=================================================================
\subsubsection*{The Quadrality Dual}
%=================================================================

\begin{figure}[ht]
	\centering	
		\resizebox{0.55\hsize}{!}{
	\includegraphics[height=6.2cm]{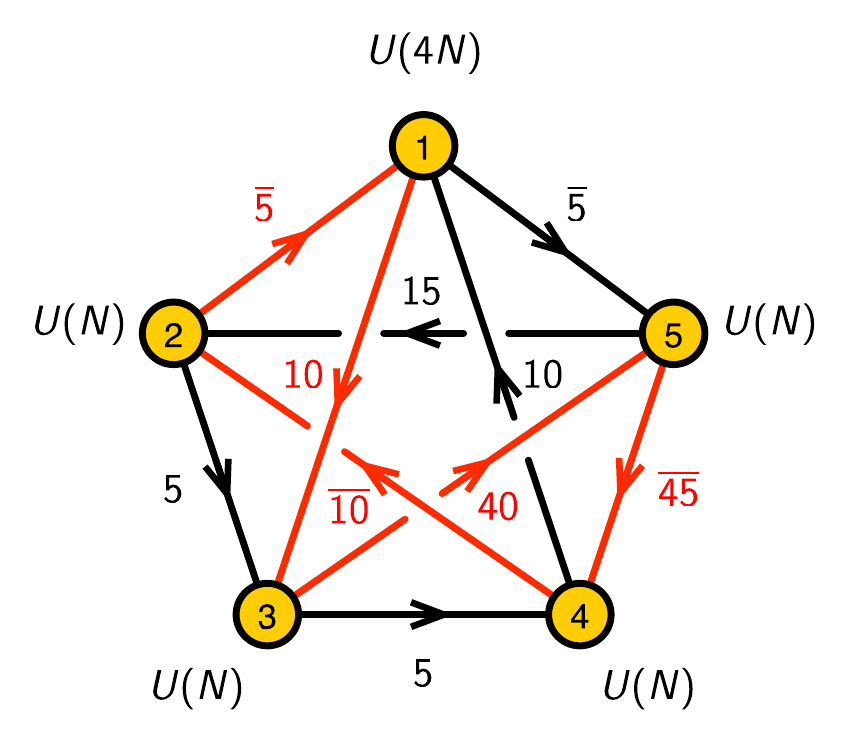}
	}
\caption{Quiver diagram for local $\mathbb{CP}^4$ obtained from the one in \fref{fig:C5Z5-quiver} by acting with quadrality on node 1.} 
	\label{quiver_C5Z5_11111}
\end{figure}

Let us now consider the theory obtained by acting with quadrality on node 1. Quadrality generates the following mesons:
\beq
\begin{array}{ccccccccc}
M_{52} & = & X_{51} \, X_{12}  & = & {\bf 5} \times {\bf 5} & = & {\bf 15} + {\bf \textcolor{blue}{10}} & \ \ \ & \mbox{(chiral)} \\[.1cm]
\mu_{54} & = & X_{51} \, \Lambda_{14} & = & {\bf 5} \times {\bf \overline{10}} & = & {\bf \textcolor{blue}{\overline{5}}} + {\bf \overline{45}} & \ \ \ & \mbox{(Fermi)} \\[.1cm]
\mu_{35} & = & \Lambda_{31} \, \overline{X}_{51} & = & {\bf \overline{10}} \times {\bf \overline{5}} & = & {\bf \textcolor{blue}{10}} + {\bf 40} & \ \ \ & \mbox{(Fermi)} 
\end{array}
\eeq
where we express them as composites of the fields in the original theory and indicate in blue the fields that become massive and are integrated out. The masses for fields in the first two lines are $J$-terms, while the one for the last line is an $H$-term. The resulting quiver diagram is shown in \fref{quiver_C5Z5_11111}. We do not write the $J$- and $H$-terms explicitly; they can be easily determined from the quiver and global symmetry. This theory is just one in an infinite web of quadrality dual theories, analogous to the web of $4d$ Seiberg dual theories on D3-branes probing local $\mathbb{CP}^2$ \cite{Cachazo:2001sg,Franco:2003ja}.

%=================================================================
%=================================================================

Chiral fields can be labeled using fundamental and antifundamental $SU(5)$ indices as follows
\beq
\phi_{52}^{ij}: {\bf 15} \qquad \phi_{23}^{k}: {\bf 5} \qquad \phi_{34}^{l}: {\bf 5} \qquad \phi_{41}^{mn}: {\bf 10} \qquad (\phi_{15})_{p}: {\bf \bar{5}}
\label{ess100}
\eeq
We restrict to $i\leq j$ and $m<n$ to match the ${\bf 15}$ (symmetric) and ${\bf 10}$ (antisymmetric) representations, respectively. 

Once again, let us focus on the case of a single D(-1)-brane, namely $N=1$. Proceeding as before, we construct gauge invariant operators using chiral fields, which transform under $SU(5)$ as follows
{\small
\beq
\begin{array}{cl}
\phi_{52}^{ij}\, \phi_{23}^{k}\, \phi_{34}^{l}\, \phi_{41}^{mn} & \hspace{-.13cm}(\phi_{15})_{p} =  {\bf 15} \times {\bf 5} \times {\bf 5} \times {\bf 10} \times {\bf \bar{5}} \\[.1cm]
& =  {\bf 1}+6({\bf 24})+7({\bf 75})+10({\bf 126})+2({\bf \overline{126}})+{\bf \underline{126}'}+7({\bf 175'})+{\bf \overline{175}'}
\\ & + 3({\bf 200})+6({\bf 224})+{\bf \overline{700}'}+4({\bf 1024})+3({\bf 1050'})+2({\bf 1701})+{\bf 1750} \, .
\end{array}
\label{SU(5)_reps_GIOs_dual}
\eeq}
where the $U(4)$ color indices at node 1 are properly contracted. For brevity, we do not provide the Dynkin labels of the representations.

The next step is to determine which of the representations in \eref{SU(5)_reps_GIOs_dual} survive once $J$-terms are taken into account. This can be conveniently done in steps, as we explain below. The key idea is the following. In this theory, all $J$-terms correspond to cubic loops in the quiver, involving one Fermi and two chiral fields. This means that the $J$-terms are quadratic in chiral fields. For each Fermi field $\Lambda_R$ transforming in some representation $R$ of $SU(5)$, the condition $J_{\Lambda_R}=0$ sets the $\overline{R}$ representation in the corresponding product of scalar fields to zero. The $J$-terms for the five types of Fermi fields in the theory end up eliminating several of the operators in \eref{SU(5)_reps_GIOs_dual}. 

It is instructive to discuss in detail how to obtain the generators. First consider the product
\beq
\phi_{34}^{i} \, \phi_{41}^{jk} = {\bf 5} \times {\bf 10}= {\bf \overline{10}}+{\bf \overline{40}} \, .
\eeq
These two chiral fields, appropriately contracted, form $J_{\Lambda_{13}}$. Since $\Lambda_{13}$ transforms in the ${\bf 10}$, we know that the ${\bf \overline{10}}$ is eliminated from the product so we are left with $\phi_{34}^{i} \, \phi_{41}^{jk}|_{J=0}={\bf \overline{40}}$. Taking the product with $(\phi_{15})_l$, we then have 
\beq
 \phi_{34}^{i} \, \phi_{41}^{jk} \, (\phi_{15})_l |_{J=0}\subseteq {\bf \overline{40}} \times {\bf \bar{5}} = {\bf 10} + {\bf 15} + {\bf 175} \, ,
\label{product_34_41_15}
\eeq
where the inclusion sign indicates that we still have not considered all the relevant $J$-terms. $J_{\Lambda_{54}}=0$ removes the $\bf{45}$ in the product $\phi_{41}^{jk} \, (\phi_{15})_l$ which, in turn, leaves us with
\beq
\phi_{34}^{i} \, \phi_{41}^{jk} \, (\phi_{15})_l |_{J=0} = {\bf 15}=[2,0,0,0] \, .
\label{product_34_41_15}
\eeq
This can be conveniently written as
\beal{ess102}
M_{(A)}^{ij}=\phi_{34}^{i} \, \phi_{41}^{jm} \, (\phi_{15})_m = \phi_{34}^{i} Z^{j} \, ,~
\label{M_A}
\eea
with $i$ and $j$ symmetrized and $Z^j =\phi_{41}^{jm} (\phi_{15})_m$. 

Next, let us consider the product
\beal{ess103}
\phi_{52}^{ij} \, \phi_{23}^{k} = {\bf 15} \times {\bf 5}={\bf \overline{35}}+{\bf \overline{40}} \, .~
\eea
$J_{\Lambda_{35}}=0$ eliminates the ${\bf \overline{40}}$, resulting in
\beq
M_{(B)}^{ijk}=\phi_{52}^{ij} \, \phi_{23}^{k} |_{J=0} = {\bf \overline{35}}=[3,0,0] \, ,
\label{M_B}
\eeq
which is symmetric in the three indices.

Combining \eref{M_A} and \eref{M_B}, we conclude the surviving gauge invariants are 
\beal{ess101}
\widetilde{M}^{m_1 m_2 m_3 m_4 m_5}_s = M_{(A)}^{m_1 m_2} M_{(B)}^{m_3 m_4 m_5} =
\phi_{52}^{m_1 m_2} \, \phi_{23}^{m_3} \, \phi_{34}^{m_4} \, Z^{m_5}  \, ,~
\eea
which transform in the ${\bf 126'}=[5,0,0,0]$ representation due to the symmetrization of all indices that follows from the remaining $J$-terms. This is in full agreement with the generators of the original theory. Furthermore, it is possible to verify explicitly that the generators satisfy the same 7000 quadratic relations of the original theory \eref{es103}. 

In summary, with the matching of generators and their first order relations, we conclude that the chiral rings of the two dual theories reproduce the coordinate ring of $\mathbb{C}^5/\mathbb{Z}_5$.

%=================================================================
\subsection{Local $(\mathbb{CP}^1)^4$}
%=================================================================

The matrix models on D(-1)-branes probing toric CY 5-folds have, in general, an infinite number of dual phases connected by quadrality. Some of these phases are described by the brane hyperbrick models discussed in section \sref{section_D-brane_probes}, equivalently by periodic quivers on $T^4$. We refer to them as {\it toric phases}. This is a straightforward generalization of the concept of toric phases in $4d$ and $2d$.

A {\it toric node} is a node in a toric phase whose dualization also results in a toric phase.\footnote{Notice that it is possible for a toric phase not to have any toric node.}  It is natural to ask what is the structure of a minimal, i.e. with a minimum number of fields, toric node for $\mathcal{N}=1$ matrix models.\footnote{We emphasize minimality because we expect multiple possibilities for toric nodes. While toric nodes have a unique structure in $4d$ theories, this is no longer the case in $2d$. It is reasonable to anticipate a similar behavior in $0d$.}

In every dimension, minimal toric nodes involve two bifundamental fields of each possible type. In the corresponding periodic quivers, the two fields in each of these pairs emanate from the toric node in opposite directions. In $4d$, such a node involves two incoming and two outgoing chirals. This configuration maps to a square face in the brane tiling \cite{Franco:2005rj}. Similarly, a minimal toric node in $2d$ has two incoming chirals, two outgoing chirals and two Fermis. This translates into a cube in the brane brick model \cite{Franco:2016nwv}. Finally, a minimal toric node in $0d$ contains two incoming chirals, two outgoing chirals, two incoming Fermis and two outgoing Fermis. It maps to a hypercube in the brane hyperbrick model.

Can we find a relatively simple toric CY 5-fold that: a) contains at least a minimal toric node and b) has more than one toric phase? For $4d$ gauge theories, a standard example of a CY$_3$ with these properties is local $\mathbb{CP}^1\times \mathbb{CP}^1$, also known as $F_0$ \cite{Franco:2005rj}. The analogous CY$_4$ for $2d$ theories is local $\mathbb{CP}^1\times \mathbb{CP}^1\times \mathbb{CP}^1$, i.e. $Q^{1,1,1}/\mathbb{Z}_2$ \cite{Franco:2016nwv}. It is then natural to conjecture that local $\mathbb{CP}^1\times \mathbb{CP}^1\times \mathbb{CP}^1\times \mathbb{CP}^1$ does the job for matrix models. Below we present some evidence supporting this claim. 

\begin{figure}[H]
	\centering
	\includegraphics[height=5.2cm]{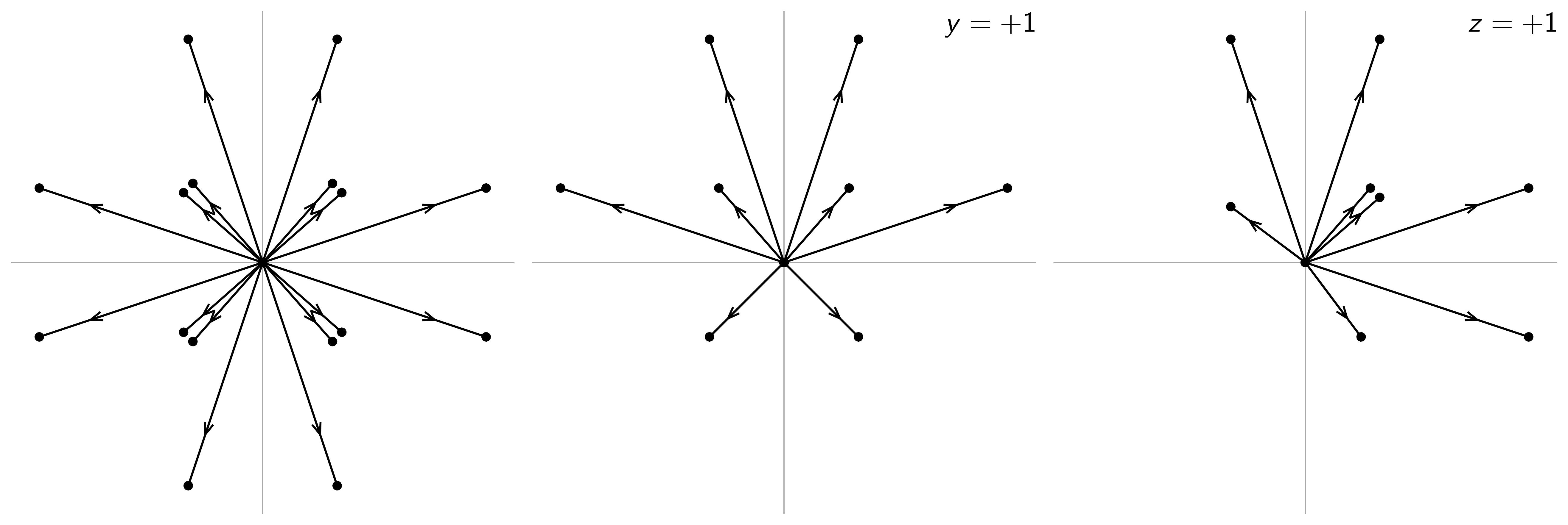}
\caption{Vanishing paths on the $W$-plane for a candidate phase of the local $(\mathbb{CP}^1)^4$ model.}
	\label{fig:P14-phase-B}
\end{figure}

Consider the following choice of coefficients in the Newton polynomial of $(\mathbb{CP}^1)^4$:
\begin{align}
P(x,y,z,w) = \left(x+\frac{1}{x}\right) + i \left(y+\frac{1}{y}\right) + 0.9(1+i) \left(z+\frac{1}{z}\right) + 0.9(-1+i)\left(w+\frac{1}{w}\right) 
\,.
\end{align}
Following the mirror symmetry analysis reviewed in section \sref{section_mirror_general}, \fref{fig:P14-phase-B}.a shows the resulting 16 critical values on the $W$-plane and the associated vanishing paths. Focusing on the $y=+1$ subset, shown in \fref{fig:P14-phase-B}.b, we recognize the configuration of vanishing paths for phase B of the $Q^{1,1,1}/\mathbb{Z}_2$ brane brick model \cite{Franco:2016nwv}.\footnote{Here we are making a comparison to a CY 4-fold. We say that the configuration of vanishing paths of the CY$_5$ is equal to the one of the CY$_4$, when they coincide on the $W$-plane.} This is a phase that indeed contains minimal toric nodes, i.e. cubic brane bricks. By symmetry, the $y=-1$ and $x=\pm 1$ subsets give rise to almost identical configurations. The $z=+1$ vanishing paths are shown in \fref{fig:P14-phase-B}.c. They also correspond to phase B of $Q^{1,1,1}/\mathbb{Z}_2$. Due to symmetry, the same is true for $z=-1$ and $w=\pm 1$. These observations, combined, suggest that it is very plausible that the configuration in \fref{fig:P14-phase-B} corresponds to a simple phase with minimal toric nodes. A detailed study of the matrix model(s) associated to this geometry would be extremely interesting. It is however beyond the scope of this paper and we leave it for future work.

%=================================================================
%=================================================================

%======================================================================
\section{Conclusions}
%======================================================================

\label{section_conclusions}

We introduced quadrality, a new order 4 duality that applies to $\mathcal{N}=1$ supersymmetric gauged matrix models. Our proposal follows naturally from mirror symmetry, which provides a unified framework that puts $4d$ Seiberg duality, $2d$ GGP triality and quadrality on an equal footing. We expect that quadrality is not restricted to theories with a D-brane realization and holds for general $\mathcal{N}=1$ matrix models. We performed various checks of the proposal, including the matching of: global symmetries, abelian flavor anomalies, deformations and the chiral ring.  The chiral ring was computed in detail for a pair of quadrality dual theories on D(-1)-branes probing local $\mathbb{CP}^4$, for which we showed not only that it is the same in the two theories but that it reproduces the coordinate ring of the CY$_5$ singularity.

We also initiated the study of various aspects of the matrix models that arise on the worldvolume of D(-1)-branes probing toric CY 5-folds.

There are various natural directions for future investigation. First, dualities are powerful tools for elucidating the dynamics of quantum field theories in different dimensions. The application of Seiberg duality to map the phase space of $4d$ $\mathcal{N}=1$ SQCD is a prime example. It would be interesting to study what quadrality can teach us about the dynamics of matrix models.
Below we discuss two additional open questions.

%======================================================================
\subsubsection*{Evaluating the Integral} 
%======================================================================

Conceptually, the most explicit way to verify quadrality would be to compute the matrix integrals for the four dual theories and show that they are equal. This would be analogous to the computation of the elliptic genus to verify $2d$ triality \cite{Gadde:2013lxa,Benini:2016qnm}. 

For triality the three dual theories flow to the same SCFT in the IR \cite{Gadde:2014ppa}. The elliptic genus is invariant under the RG flow, making it possible to probe the SCFT from the UV gauge theory. In $0d$, the usual notion of RG flow does not exist, so determining the conditions under which the matrix integrals should agree becomes more subtle.

The $2d$ elliptic genus can be refined by turning on flavor fugacities. In the case of non-compact target spaces, the fugacities regulate divergent contributions. In $0d$, since there are no background gauge fields, it is not clear how to turn on fugacities. 

Due to these issues, evaluating the integrals and comparing between different quadrality phases is not straightforward. We hope to revisit these questions in future work.

%======================================================================
\subsubsection*{M-Theory Lift}
%======================================================================

Mirror symmetry relates D(-1)-branes probing a CY 5-fold to Euclidean D4-branes wrapping 5-cycles in the mirror CY$_5$. It would be interesting to determine the M-theory lift of this configuration. The ED4-branes become EM5-branes wrapping the original 5-cycles times the M-theory circle. Wick-rotating and decompactifying the M-theory circle, we arrive at an M-theory configuration with physical M5-branes wrapping 5-cycles. The result is a supersymmetric quantum mechanics. The original $0d$ matrix model can be reinterpreted as the dimensional reduction of the M-theory quantum mechanics. 

The situation is somewhat analogous, but not equivalent, to the relation between the D-instanton matrix model and D0 quantum mechanics. In particular, the D(-1) matrix model contains information on the Witten index of D0 quantum mechanics \cite{Green:1998yf}. In the D0/D(-1) connection, an explicit Lagrangian description is available on both sides. In contrast, in the relation between the IIA matrix model and the M-theory quantum mechanics we are considering, the Lagrangian is only known for the former but not for the latter. The precise nature of the quantum mechanics of wrapped M5-branes is an interesting open problem. 

More generally, it would also be interesting to determine whether some new duality for supersymmetric quantum mechanics can be inferred from quadrality.

%======================================================================

%======================================================================  
\acknowledgments
%======================================================================  

We would like to thank P. Putrov, M. Romo, N. Seiberg, E. Witten and S.-T. Yau for useful and enjoyable discussions. We are also grateful to D. Ghim for collaboration on related topics. We gratefully acknowledge support from the Simons Center for Geometry and Physics, Stony Brook University, where some of the research for this paper was performed during the 2016 Simons Summer Workshop. The work of S. F. is supported by the U.S. National Science Foundation grant PHY-1518967 and by a PSC-CUNY award. The work of S. L. was supported by Samsung Science and Technology Foundation under Project Number SSTBA1402-08. The work of S. L. was also performed in part at the Institute for Advanced Study supported by the IBM Einstein Fellowship of the Institute for Advanced Study, and at the Aspen Center for Physics supported by National Science Foundation grant PHY-1066293. The work of R.-K. S. is supported by the ERC STG grant 639220 ``Curved SUSY''. The work of C.V. is supported in part by NSF grant PHY-1067976.

\appendix

%======================================================================
\section{The Hilbert Series for $\mathbb{C}^3/\mathbb{Z}_3$, $\mathbb{C}^4/\mathbb{Z}_4$ and $\mathbb{C}^5/\mathbb{Z}_5$ \label{sapphs}}
%======================================================================

The Hilbert series \cite{Benvenuti:2006qr, Feng:2007ur} is a powerful tool for enumerating operators in a chiral ring and for studying its geometric structure. It is formally defined in algebraic geometry as the generating function 
\beal{esapp1}
g(t;R) = \sum_{n=0}^{\infty} \text{dim}(R_n) t^n \, ,~
\eea
where $R$ is an algebraic quotient ring and $R_n$ is a component of $R$ of degree $n\in \mathbb{N}$. The fugacity $t$ counts the degree of the component $R_n$. For multi-graded rings with components $R_{\vec{n}}$ and grading $\vec{n}=(n_1,\dots,n_k)$, the Hilbert series takes the form $g(t_1,\dots,t_k;R) = \sum_{n=0}^{\infty} \text{dim}(R_{\vec{n}}) t_1^{n_1}  \dots t_k^{n_k}$, where $t_1,\dots,t_k$ are the fugacities of the grading.

\begin{figure}[htbp]
	\centering	
	\resizebox{0.7\hsize}{!}{
	\includegraphics[height=5.5cm]{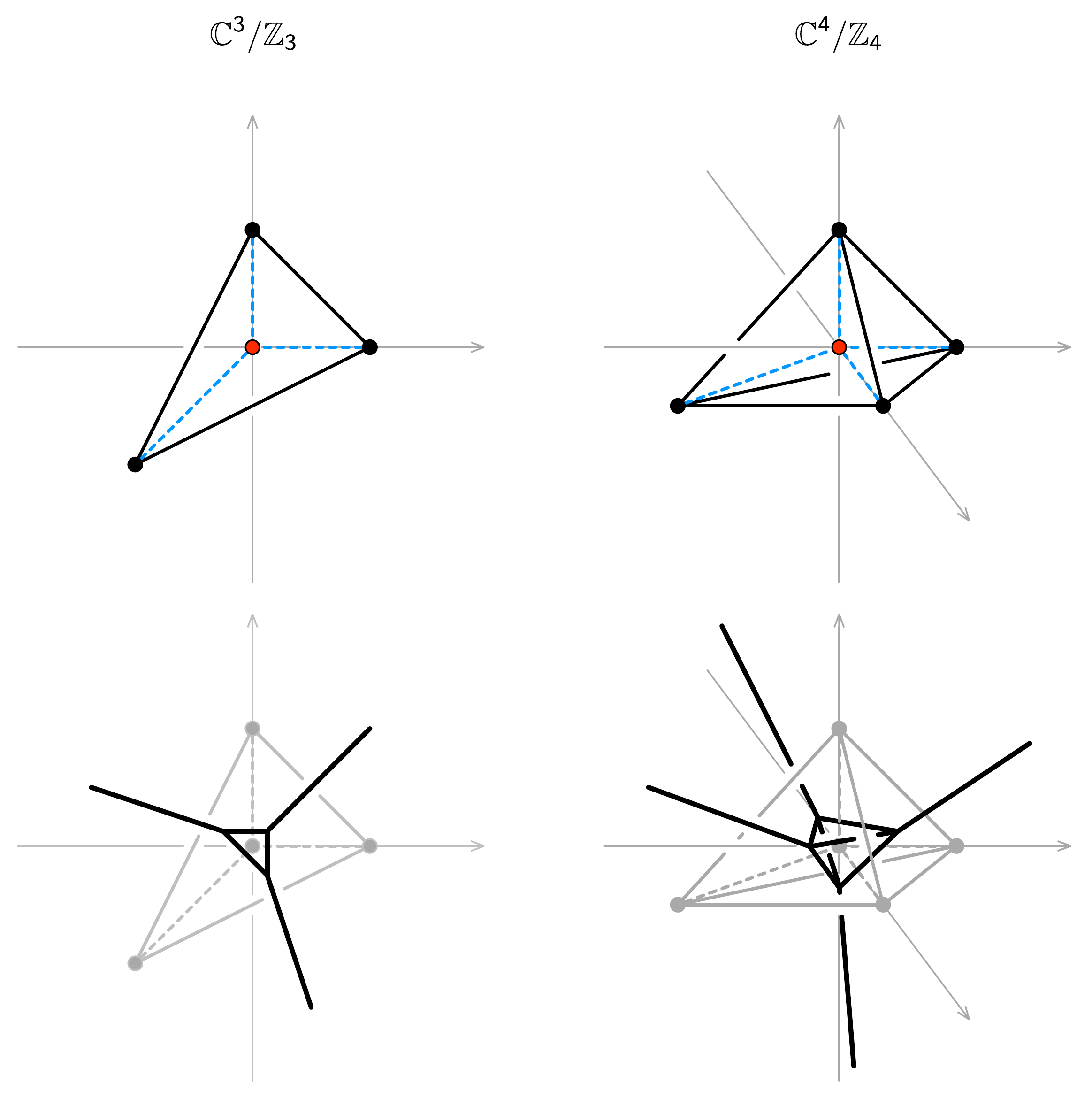}
	}
\caption{The toric diagrams and dual web diagrams for $\mathbb{C}^3/\mathbb{Z}_3$ and $\mathbb{C}^4/\mathbb{Z}_4$.} 
	\label{fpwebs}
\end{figure}

%=================================================================
\paragraph{Hilbert Series from Toric Geometry.} 
%=================================================================
When the chiral ring is a toric variety, its Hilbert series can be derived directly from the toric diagram \cite{Benvenuti:2006qr,Martelli:2006yb}. For a toric CY $n$-fold, the toric diagram is an $n-1$ dimensional convex polytope that admits at least one triangulation in terms of $(n-1)$-simplices. From a triangulation, one can construct a dual web diagram. For Calabi-Yau 3-folds, these are the so-called $(p,q)$-webs \cite{Aharony:1997ju,Aharony:1997bh,Leung:1997tw}. \fref{fpwebs} shows the toric diagrams and dual web diagrams for $\mathbb{C}^3/\mathbb{Z}_3$ and $\mathbb{C}^4/\mathbb{Z}_4$.

The Hilbert series of the toric variety $X$ then can be defined from a triangulation of the toric diagram as follows
\beal{esapp10}
g(t_1,\dots,t_n; X) = \sum_{i=1}^{r} \prod_{j=1}^{n} (1- \vec{t}^{~\vec{v}(i,j)}) \, ,~
\eea
where $\vec{t}^{~\vec{v}(i,j)}= \prod_{a}^{n} t_a^{v_a(i,j)}$. The index $i=1,\dots,r$ runs over the simplices making up the triangulation while $j=1,\dots,n$ runs over the faces of each simplex. The vector $\vec{v}(i,j)$ is the $n$-dimensional outer normal to the face of the fan associated to face $j$ of simplex $i$.\footnote{Notice that for a CY $n$-fold, these normal vectors are $n$-dimensional, while we said that the toric and web diagrams are $(n-1)$-dimensional. More precisely, the toric diagram of a CY $n$-fold lives on a hyperplane in $n$-dimensions at distance 1 from the origin. This fact allows a trivial projection to $(n-1)$-dimension. In the examples below, we reincorporate the $n^{th}$ coordinate in order to determine the normal vectors.}

%=================================================================
%=================================================================

%=================================================================
\paragraph{Hilbert Series for $\mathbb{C}^3/\mathbb{Z}_3$, $\mathbb{C}^4/\mathbb{Z}_4$, and $\mathbb{C}^5/\mathbb{Z}_5$.} 
%=================================================================
The coordinates for the points in the toric diagrams for these three geometries are
\beal{esapp15}
\mathbb{C}^{3}/\mathbb{Z}_3 &~:~&
(1,0,1),(0,1,1),(-1,-1,1),(0,0,1)
\nn\\
\mathbb{C}^{4}/\mathbb{Z}_4 &~:~&
(1,0,0,1),(0,1,0,1),(0,0,1,1),(-1,-1,-1,1),(0,0,0,1)
\nn\\
\mathbb{C}^{4}/\mathbb{Z}_4 &~:~&
(1,0,0,0,1),(0,1,0,0,1),(0,0,1,0,1),(0,0,0,1,1),(-1,-1,-1,-1,1),\nn\\
&&(0,0,0,0,1)
\nn\\
\eea
Notice that we have included the $n^{th}$ coordinate, which will become important in the discussion that follows. Labeling the points in the toric diagrams from $1,\dots,n+1$ in the order they are listed in \eref{esapp15}, then the unique triangulations of these diagrams can be summarized in terms of the points in the toric diagram as follows
\beal{esapp16}
\mathbb{C}^{3}/\mathbb{Z}_3 &~:~&
\{\{1,2,4\},\{1,3,4\}\,\{2,3,4\}\} 
\nn\\
\mathbb{C}^{4}/\mathbb{Z}_4 &~:~&
\{\{1, 2, 3, 5\}, \{1, 2, 4, 5\}, \{1, 3, 4, 5\}, \{2, 3, 4, 5\}\}
\nn\\
\mathbb{C}^{4}/\mathbb{Z}_4 &~:~&
\{\{1, 2, 3, 4, 6\}, \{1, 2, 3, 5, 6\}, \{1, 2, 4, 5, 6\}, \{1, 3, 4, 5, 6\}, \{2, 3, 4, 5, 6\}\}
\nn\\
\eea
The corresponding Hilbert series are
\beal{esapp17}
g(t;\mathbb{C}^3/\mathbb{Z}_3) &=&
\frac{1 + 7 t + t^2}{(1 - t)^3} \, ,~
\nn\\
g(t;\mathbb{C}^4/\mathbb{Z}_4) &=&
\frac{1 + 31 t + 31 t^2 + t^3}{(1 - t)^4} \, ,~
\nn\\
g(t;\mathbb{C}^5/\mathbb{Z}_5) &=&
\frac{1 + 121 t + 381 t^2 + 121 t^3 + t^4}{(1 - t)^5} \, .~
\eea
The plethystic logarithms \cite{Benvenuti:2006qr,Feng:2007ur} of the Hilbert series are 
\beal{esapp18}
\text{PL}[g(t;\mathbb{C}^3/\mathbb{Z}_3)] &=&  10 t - 27 t^2 + 105 t^3 - 540 t^4 + \dots \, ,~
\nn\\
\text{PL}[g(t;\mathbb{C}^4/\mathbb{Z}_4)] &=&  35 t - 465 t^2 + 8960 t^3 - 201376 t^4 + \dots \, ,~
\nn\\
\text{PL}[g(t;\mathbb{C}^5/\mathbb{Z}_5)] &=&  126 t - 7000 t^2 + 544500 t^3 - 48095250 t^4 + \dots \, .~
\eea
Note that none of these plethystic logarithms has a finite expansion, indicating that the three toric varieties are not complete intersections. Furthermore, the coefficients in \eref{esapp18} are sums of dimensions of irreducible representations of $SU(3)$, $SU(4)$ and $SU(5)$, respectively. 

This computation confirms that the chiral ring discussed in \sref{scoordring} indeed corresponds to the coordinate ring of $\mathbb{C}^5/\mathbb{Z}_5$. The first term in the plethystic logarithm corresponds to the 126 generators of the toric variety transforming in the $[5,0,0,0]$ representation of $SU(5)$. The second term in the expansion indicates that these 126 generators satisfy 7000 quadratic relations, which transform in the $[6,2,0,0]+[2,4,0,0]$ representations of $SU(5)$.

%======================================================================
\section{Review: $\mathbb{C}^3/\mathbb{Z}_3$ \label{sappc3z3}}
%======================================================================

\begin{figure}[htbp]
	\centering	
	\resizebox{0.35\hsize}{!}{
	\includegraphics[height=5.5cm]{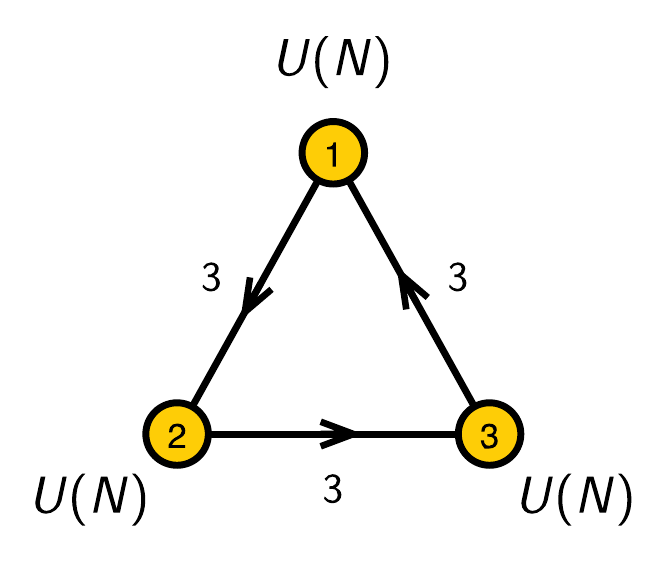}
	}
\caption{Quiver diagram for $N$ D3-branes probing $\mathbb{C}^3/\mathbb{Z}_3$.}
	\label{fc3z3quiver}
\end{figure}

While the D-brane constructions of this paper focus on D(-1)-branes probing toric CY 5-folds, it is enlightening to review the case of D3-branes probing $\mathbb{C}^3/\mathbb{Z}_3$ in further detail. This example is useful because it exhibits many of the concepts that apply to the CY$_5$ case in a considerably simpler context. For $N$ D3-branes, the worldvolume theory is a $4d$ $\mathcal{N}=1$ supersymmetric gauge theory with the quiver diagram shown in \fref{fc3z3quiver} and superpotential
\beal{esapp30}
W= \epsilon_{ijk} \phi_{12}^i \phi_{23}^j \phi_{31}^k \, .~
\eea
The global symmetry of the theory is $SU(3)\times U(1)_R$. 

%=================================================================
%=================================================================

The Hilbert series for a single D3-brane has been computed in \cite{Hanany:2012hi}. It takes the form
\beq
g(t,x_1,x_2; \mathbb{C}^3/\mathbb{Z}_3) = \sum_{n=0}^{\infty} [3n,0] t^n \, ,~
\label{esapp31}
\eeq
where $[3n,0]$ is the character of the $SU(3)$ representation, with the entries being Dynkin labels of the representation. $t$ is the fugacity for the $U(1)_R$ symmetry. The plethystic logarithm is
\beal{esapp31}
\text{PL}[g(t,x_1,x_2; \mathbb{C}^3/\mathbb{Z}_3)] =  [3,0] t - [2,2] t^2 + ([1,1]+[1,4]+[2,2]+[4,1]) t^3 + \dots \, .~
\nn\\
\eea
This matches the result obtained from toric geometry in \eref{esapp18}. There are 10 generators which transform in the $[3,0]$ representation of $SU(3)$ satisfying 27 relations transforming in the $[2,2]$ representation of $SU(3)$. The vacuum moduli space is not a complete intersection and for the abelian theory it is precisely $\mathbb{C}^3/\mathbb{Z}_3$. 

The 10 generators can be written in terms of the chiral bifundamental fields $\phi_{ab}^{i}$ as follows,
\beq
M^{ijk} =  \phi_{12}^i \phi_{23}^j \phi_{31}^k \, ,~
\label{esapp32}
\eeq
where one sets $i\leq j \leq k$ such that $M^{ijk}$ transform in the $[3,0]$ representation of $SU(3)$. This follows from the fact that the $F$-terms associated to the superpotential in \eref{esapp30},
\beal{esapp33}
\frac{\partial W}{\partial \phi_{ab}^{i}} = \epsilon_{ijk} \phi_{bc}^j \phi_{ca}^k = 0 \, .~
\eea
fully symmetrize the indices of $M^{ijk}$. 

The plethystic logarithm \eref{esapp31} indicates that there are quadratic relations between the generators $M^{ijk}$ transforming in the $[2,2]$ representation of $SU(3)$. We can identify these relations as 
\beal{esapp34}
R^{ij}_{mn} = \frac{1}{2} M^{i k_1 k_2} M^{j l_1 l_2} \epsilon_{k_1 l_1 m} \epsilon_{k_2 l_2 n} = 0\, .~
\eea

%=================================================================
\begin{figure}[htbp]
	\centering	
	\resizebox{0.7\hsize}{!}{
	\includegraphics[height=5.5cm]{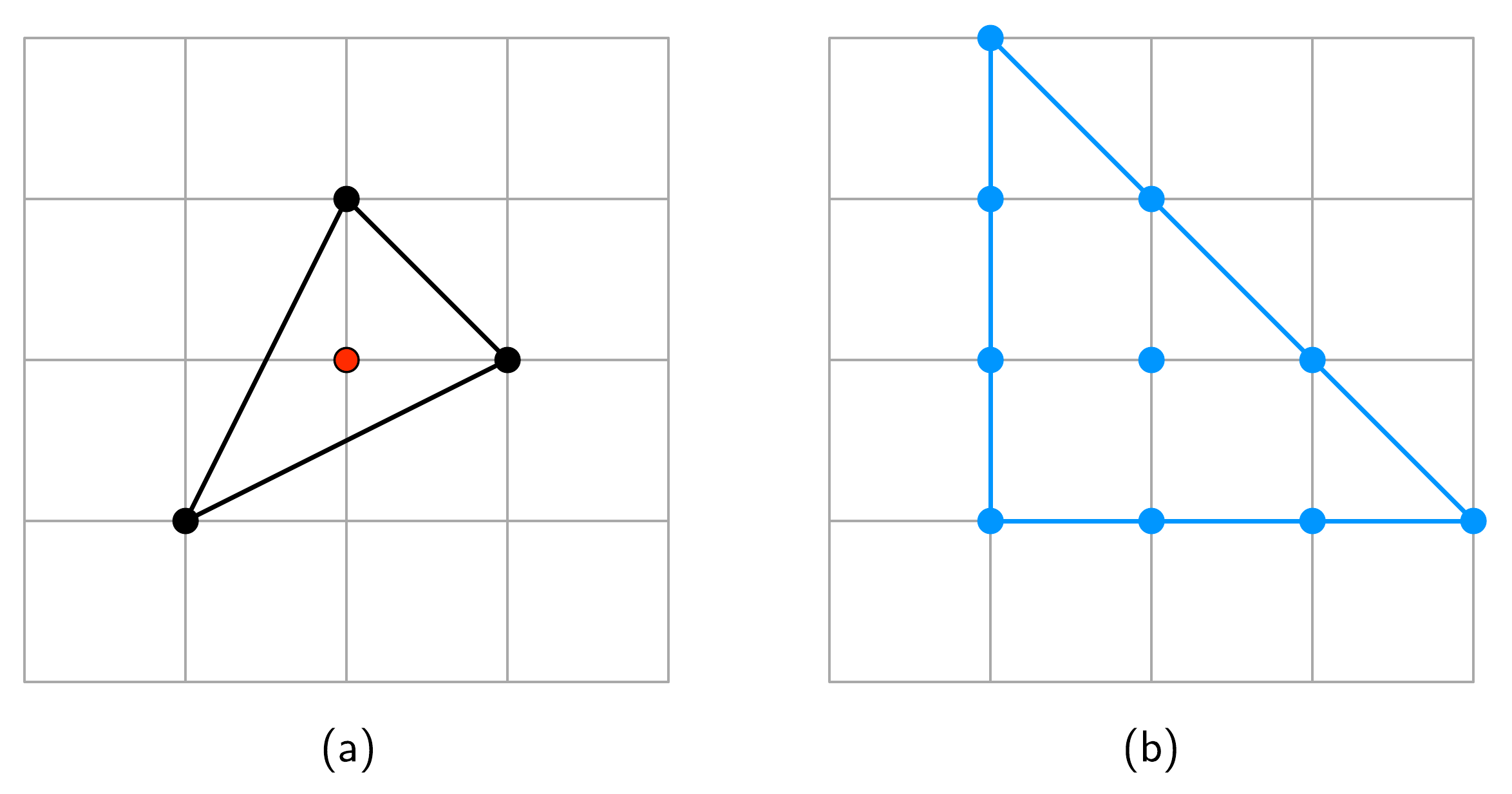}
	}
\caption{(a) The toric diagram for $\mathbb{C}^3/\mathbb{Z}_3$ and (b) the lattice of its generators. Note that the generators form the convex polygon dual to the toric diagram.}
	\label{fc3z3gen}
\end{figure}
%=================================================================

Knowing the generators and their first order relations, $\mathbb{C}^3/\mathbb{Z}_3$ can be expressed as the following quotient variety,
\beq
M^{ijk} / \langle R^{ij}_{mn} = 0 \rangle \, .~
\label{esapp35}
\eeq
The $SU(3)$ charges of the generators, which correspond to the exponents of $x_1^{n_x} x_2^{n_y}$ in the character of $[3,0]$,
\beal{esapp36}
[3,0] = \frac{x_2^3}{x_1^3}+x_1^3+\frac{x_1^2}{x_2}+\frac{x_2}{x_1^2}+\frac{x_1}{x_2^2}+\frac{x_2^2}{x_1}+x_1x_2+\frac{1}{x_1 x_2}+\frac{1}{x_2^3}+1 \, ,~
\eea
can be plotted on a $\mathbb{Z}^2$ lattice. It is possible to transform the lattice points associated to the generators such that the new coordinates are $(n_y-2n_x,-n_x-n_y)$, which is an $SL(2,\mathbb{Z})$ transformation and a rescaling.
The resulting points form a lattice triangle, which is the dual reflexive polygon of the toric diagram of $\mathbb{C}^3/\mathbb{Z}_3$ as illustrated in \fref{fc3z3gen} \cite{Hanany:2012hi}.

%======================================================================
\bibliographystyle{JHEP}
\bibliography{mybib}
%======================================================================

\end{document}